\documentclass[a4paper,12pt]{article}
\pdfoutput=1
\usepackage{epsfig,graphicx,xcolor,amsbsy,amssymb,latexsym,amsfonts,amsmath,,tcolorbox,setspace}
\usepackage{pstricks}
\usepackage{color}
\usepackage{soul}
\usepackage[numbers,square,comma, compress]{natbib}
\usepackage{placeins}
\usepackage{wrapfig}

\usepackage{eurosym}
\usepackage[a4paper]{geometry}
\usepackage{graphicx}
\usepackage{tikz}
\usepackage{xcolor}
\usepackage{amsmath,amssymb,amsfonts,pstricks,setspace}
\usepackage[enableskew]{youngtab}

\numberwithin{equation}{section}

\newcommand{\bea}{\begin{eqnarray}\displaystyle}
\newcommand{\eea}{\end{eqnarray}}

\newcommand{\figref}[1]{Fig.~\protect\ref{#1}}

\newcommand{\subg}{\mathfrak{k}}

\newcommand{\unk}{\mathbb{W}}
\newcommand{\unc}{\mathfrak{w}}

\title{
\begin{flushright}{\vspace{-2.5cm}\small LYCEN 2019-01\\}\end{flushright}
\vspace{2.3cm}
{\bf Symmetries in A-Type Little String Theories, Part I}\\[40pt]
{\Large Reduced Free Energy and Paramodular Groups}\\[45pt]}

\author{\large \textsc{Brice Bastian\footnote{\tt b.bastian@uu.nl}}\,\,\, and\, \textsc{Stefan~Hohenegger\footnote{\tt s.hohenegger@ipnl.in2p3.fr}}}
\date{}

\begin{document}

\maketitle

\begin{center}
\renewcommand{\thefootnote}{\fnsymbol{footnote}}\vspace{-0.5cm}
${}^{\footnotemark[1]\footnotemark[2]}$ Univ Lyon, Univ Claude Bernard Lyon 1, CNRS/IN2P3, IP2I Lyon,\\ UMR 5822, F-69622, Villeurbanne, France\\[0.5cm]
\renewcommand{\thefootnote}{\fnsymbol{footnote}}
${}^{\footnotemark[1]}$ Institute~for~Theoretical~Physics\\ Utrecht University, Princetonplein 5, 3584 CE Utrecht, The Netherlands\\[2.5cm]
\end{center}

\begin{abstract}
We analyse the symmetries of a class of A-type little string theories that are engineered by $N$ parallel M5-branes with M2-branes stretched between them. This paper deals with the so-called reduced free energy, which only receives contributions from the subset of the BPS states that carry the same charges under all the Cartan generators of the underlying gauge algebra. We argue (and check explicitly in a number of examples) that the former is invariant under the paramodular group $\Sigma_N\subset Sp(4,\mathbb{Q})$, which gets extended to a subgroup of $Sp(4,\mathbb{R})$ in the Nekrasov-Shatashvili-limit. This extension agrees with the observation made in \cite{Ahmed:2017hfr} that these BPS states form a symmetric orbifold CFT. Furthermore, we argue that $\Sigma_N$ (along with other symmetries) places strong constraints on the BPS counting function that governs the intersection between the M5- and M2-branes.
\end{abstract}

\newpage

\tableofcontents

\onehalfspacing

\vskip1cm

\section{Introduction}
The study of quantum theories in six dimensions as well as their compactifications to lower dimensions through various string constructions has attracted a lot of attention in recent years. Besides a lot of activity related to superconformal field theories (SCFTs), culminating in the recent classification through geometric means \cite{Heckman:2015bfa}, a lot of work has been dedicated to so-called \emph{Little String Theories} (LSTs) \cite{Seiberg:1997zk,Aharony:1999ks,Kutasov:2001uf}. The latter allow a low energy description in terms of supersymmetric gauge theories (containing only point-particle degrees of freedom), however, contain a characteristic scale, beyond which the inclusion of string-like degrees of freedom is required. These theories admit an ADE-type classification \cite{Bhardwaj:2015oru} and a particularly rich two-parameter class of models of A-type (with different amounts of supersymmetry) can be constructed from M-theory: they arise as a particular decoupling limit (which in particular decouples the gravitational sector) of a system of $N$ parallel M5-branes on a circle that probe a transverse $\mathbb{Z}_M$-orbifold singularity \cite{Haghighat:2013gba,Haghighat:2013tka,Hohenegger:2013ala,Hohenegger:2015btj}. Exploiting various dual descriptions of this M-theory setup \cite{Haghighat:2013gba,Haghighat:2013tka,Hohenegger:2015btj}, notably F-theory compactified on a particular class of toric Calabi-Yau threefolds $X_{N,M}$ with the structure of a double elliptic fibration \cite{Kanazawa:2016tnt}, has allowed to compute the full non-perturbative BPS partition function $\mathcal{Z}_{N,M}(\omega,\epsilon_{1,2})$ \cite{Hohenegger:2015btj} for these theories on $\mathbb{R}^4 \times T^2$. Here $\omega$ corresponds to a set of $NM+2$ parameters related to the M-brane setup (which correspond to K\"ahler parameters of $X_{N,M}$ from the F-theory perspective), while $\epsilon_{1,2}$ are deformation parameters that are required to render the partition function well-defined and, from the perspective of the low-energy gauge theory description, correspond to the $\Omega$-background \cite{Nekrasov:2002qd}. For generic values of $(N,M)$ this low energy theory is a six-dimensional quiver gauge theory with $M$ nodes of gauge groups $U(N)$ and matter in the bifundamental representation, which we shall denote as $[U(N)]^M$ in the following.

The partition function $\mathcal{Z}_{N,M}$ can be written in the form of a series expansion in a subset of the parameters $\omega$, which can be interpreted as an instanton sum from the perspective of the theory $[U(N)]^M$. However, this description is in general not unique: fiber-base duality of $X_{N,M}$ (or general string S-duality) suggests that the theory $[U(N)]^M$ is dual to $[U(M)]^N$, which implies the existence of an equivalent expansion of $\mathcal{Z}_{N,M}$ in terms of a different subset of $\omega$ that matches the instanton series of the latter theory. Moreover, as was recently been pointed out in a series of works \cite{Bastian:2017ing,Bastian:2017ary,Bastian:2018dfu}, there exist numerous other theories dual to $[U(N)]^M$, each of which entailing a new (but equivalent) expansion of $\mathcal{Z}_{N,M}$. More precisely, based on geometric considerations related to the extended moduli space of $X_{N,M}$ it was conjectured \cite{Hohenegger:2016yuv} that the theory $[U(N')]^{M'}$ is dual to $[U(N)]^M$ if $NM=N'M'$ and $\text{gcd}(N',M')=\text{gcd}(N,M)$. This conjecture was proven at the level of the partition function for $M=1$ in \cite{Bastian:2017ing} and, by studying the Seiberg-Witten curve related to the Calabi-Yau geometry, in \cite{Haghighat:2018gqf} for generic $(M,N)$ (however for vanishing parameters $\epsilon_{1,2}$). 

The web of dual (quiver) gauge theories is very astonishing from a purely field theoretic point of view, since the duality maps are generically non-perturbative in nature, exchanging coupling constants, gauge- and mass parameters in a rather non-trivial fashion. In \cite{Bastian:2018jlf} the question was raised, whether this web also has sizeable implications for an individual gauge theory, such as additional symmetries. Focusing on the case $\text{gcd}(N,M)=1$ (and thus working in the case $M=1$) and guided once more by the structure of the extended moduli space of $X_{N,1}$, we argued for the existence of a symmetry group of the form $\widetilde{\mathbb{G}}(N)\cong\mathbb{G}(N)\times \mathcal{S}_N$, where $\mathcal{S}_N\subset S_N$ is (a subgroup of) the Weyl group of the $U(N)$ gauge group (\emph{i.e.} the largest single gauge group factor that can be constructed from $X_{N,1}$), while
\begin{align}
\mathbb{G}(N)\cong\left\{\begin{array}{lcl}\text{Dih}_3 & \text{if} & N=1\,, \\ \text{Dih}_N & \text{if} & N=2,3\,,\\\text{Dih}_\infty & \text{if} & N\geq 4\,.\end{array}\right.\label{FirstIntroDihedral}
\end{align}
Here $\text{Dih}_\infty$ is the group freely generated by two elements of order 2 that satisfy no additional braid relation. The group $\widetilde{\mathbb{G}}(N)$ has a natural action on the single-particle single free energy $F_{N,1}(\omega,\epsilon_{1,2})$ associated with the partition function $\mathcal{Z}_{N,1}$, as $(N+2)\times (N+2)$ dimensional matrices acting linearly on the Fourier coefficients of $F_{N,1}(\omega,\epsilon_{1,2})$.

The group $\widetilde{\mathbb{G}}(N)$ is certainly not the full symmetry group of the gauge theories engineered by $X_{N,1}$. For one, it was argued in \cite{Hohenegger:2015btj} (see also \cite{Haghighat:2013gba}) that $F_{N,1}$ (when suitably expanded) has specific modular properties under two $SL(2,\mathbb{Z})$ symmetries related to the modular parameters of the double elliptic fibration structure of $X_{N,1}$. The general structure of the combined symmetry group is not obvious from this perspective (see, however, also \cite{Haghighat:2018gqf} for a recent discussion from the perspective of the Seiberg-Witten curve), however, this was argued in \cite{Bastian:2018jlf} that in a particular region in the moduli space (in which most K\"ahler parameters of $X_{N,1}$ are identified), generically a subgroup of $Sp(4,\mathbb{Z})$ is realised, which is isomorphic to $Sp(4,\mathbb{Z})$ in the case $N=1$.

This paper together with the companion paper~\cite{Companion}, continues the study of symmetries of a single gauge theory in the duality web generated by $X_{N,1}$, by analysing their action on the Fourier coefficients of $F_{N,1}$. Keeping the moduli of the theory generic, in the present work we focus on a particular subsector of the free energy, which in \cite{Ahmed:2017hfr} was called the \emph{reduced free energy} $\mathcal{F}^{N}$, and study the remaining sectors in \cite{Companion}. Concretely, the former was defined in \cite{Ahmed:2017hfr} as follows\footnote{In \cite{Ahmed:2017hfr} the full free energy (as defined in eq.~(\ref{AbstDefOrbSect}) was discussed). In the present work, we shall exclusively focus on the free energy constructed by counting irreducible single particle states (see eq.~(\ref{PletFreeEnergy})), which is obtained from (\ref{AbstDefOrbSect}) by replacing the $\ln$ with the plethystic logarithm \cite{Feng:2007ur}. The main reason for this choice is that it is computationally more accessible, while still retaining very similar properties as the full free energy. In fact, all of the results obtained in this paper can directly be carried over to the latter.}
\begin{align}
\mathcal{F}^{N}(R,S,\rho,\epsilon_{1,2})=\sum_{k,n\geq0}Q_\rho^k Q_R^n\prod_{i=1}^N\oint\frac{dQ_{\widehat{a}_i} }{Q_{\widehat{a}_i}^{k+1}}\,\ln\mathcal{Z}_{N}(\omega,\epsilon_{1,2})\,, \label{AbstDefOrbSect}
\end{align}
where $Q_{\widehat{a}_i}$ (for $i=1,\ldots,N$) are the (exponentiated) K\"ahler parameters associated with the gauge structure of (affine) $\widehat{\mathfrak{a}}_{N-1}$ and the contour integrals are understood locally around the point $Q_{\widehat{a}_i}=0$. The contributions (\ref{AbstDefOrbSect}) are not very sensitive to the details of the gauge group since it only capture states that carry the same charges (captured by the summation over $k$ in (\ref{AbstDefOrbSect})) under all the generators of the Cartan subalgebra of $\mathfrak{a}_{N-1}$. In \cite{Ahmed:2017hfr}, this sector was selected, since it only counts monopole charges\footnote{This point of view uses a dual description of the LSTs, in terms of monopole strings in 5-dimensions.} of the type $(k,\ldots,k)$ for $k\in\mathbb{N}$. The reduced free energy (\ref{AbstDefOrbSect}) depends (besides $\epsilon_{1,2}$) on 3 remaining parameters $(R,S,\rho)$, out of which one can interpret $Q_\rho=\prod_{i=1}^NQ_{\widehat{a}_i}$. It was observed in \cite{Ahmed:2017hfr}  that $\mathcal{F}^{N}$ carries a particular Hecke structure, which is the  hallmark of a symmetric orbifold CFT \cite{Dijkgraaf:1996xw}. It was therefore conjectured, that the BPS states of the M5-brane system, realised by M2-branes stretched between the former, that contribute to $\mathcal{F}^{N}$ form such a conformal field theory.

Our analysis is based on the improved understanding provided by the symmetry group $\widetilde{\mathbb{G}}(N)$, which not only provides us with a new (more group theoretic) perspective on the symmetries of $\mathcal{F}^{N}(R,S,\rho,\epsilon_{1,2})$ but also allows us to make statements of the local intersection of the M5-branes with the M2-branes (dubbed the M-string in \cite{Haghighat:2013gba}). Concretely, we find two main results in this work
\begin{itemize}
\item For generic values of the parameters $\epsilon_{1,2}$, the reduced free energy $\mathcal{F}^{N}(R,S,\rho,\epsilon_{1,2})$ is covariant under the paramodular group $\Sigma_N\subset Sp(4,\mathbb{Q})$, which acts in the canonical manner on the period matrix $\Omega_N=\left(\begin{array}{cc} R & S \\ S & \rho/N\end{array}\right)$. This action in particular unifies the modular transformations on $\rho$ and $R$ respectively, as well as the group $\mathbb{G}(N)$ in (\ref{FirstIntroDihedral}). In the Nekrasov-Shatashvili (NS) limit \cite{Nekrasov:2009rc} (which essentially corresponds to $\epsilon_2\to 0$), we provide evidence that the paramodular group $\Sigma_N$ is enhanced by a single generator to $\Sigma_N^*\subset Sp(4,\mathbb{R})$, where the additional symmetry acts as the exchange $R\longleftrightarrow \rho$. The latter acts at the level of the Fourier coefficients of $\mathcal{F}^{N}$, thus going beyond the S-duality that relates $\mathcal{Z}_{N,1}$ to $\mathcal{Z}_{1,N}$. The presence of the symmetry group $\Sigma_N^*$ would provide further strong evidence for the Hecke structure described in \cite{Ahmed:2017hfr}, as its presence is directly implied by the latter according to \cite{Gritsenko1995}.

\item In \cite{Hohenegger:2015btj} it was observed that the BPS partition function of $N$ M5-branes with M2-branes stretched between them can be recovered from the partition function of $N-1$ M5-brane, provided one of the branes has only a single M2-brane ending on it on both sides. A crucial role in this equivalence is played by a modular object $W(R,S,\epsilon_{1,2})$, which in a sense governs the intersection between the M5- and the M2-brane. In the current paper we argue, that the modular symmetries as well as $\mathbb{G}(N)$, along with certain limits that enhance the supersymmetry content of the brane setup, pose strong constraints on $W(R,S,\epsilon_{1,2})$. In fact, considering an expansion up to order $O(\epsilon_{1,2}^7)$ we show that the latter is uniquely fixed upon also imposing an appropriate normalisation condition on the free energy. 
\end{itemize}

This paper is organised as follows: Section~\ref{Sect:PartitionFunctionSyms} is dedicated to a review of the partition function $\mathcal{Z}_{N,1}$ as well as the associated (reduced) free energy along with some of their symmetries that have been previously discussed in the literature. Section~\ref{Sect:ParamodularEnhancement} provides evidence for the covariance of $\mathcal{F}^N(R,S,\rho,\epsilon_{1,2})$ under the paramodular group $\Sigma_N$ and its enhancement in the NS-limit $\epsilon_2\to 0$. We also clarify the role that $\mathbb{G}(N)$ takes in $\Sigma_N$. Section~\ref{Sect:Reconstruction} discusses constraints on the function $W(R,S,\epsilon_{1,2})$ imposed by various symmetries of the free energy. Finally, Section~\ref{Sect:Conclusions} contains our conclusions and a short outlook on the companion paper \cite{Companion}. A short review on (quasi-)modular objects and the paramodular group, as well a selection of explicit expressions of Fourier coefficients of $\mathcal{F}^N(R,S,\rho,\epsilon_{1,2})$ that have been deemed too long to be included in the main body of the paper have been relegated to three appendices.


\section{Little String Partition Function and its Symmetries}\label{Sect:PartitionFunctionSyms}
\subsection{Web Diagram and K\"ahler Parameters}
\begin{wrapfigure}{l}{0.52\textwidth}
\begin{center}
\vspace{-0.5cm}
\scalebox{0.7}{\parbox{11.2cm}{\begin{tikzpicture}[scale = 1.50]
\draw[ultra thick] (-1,0) -- (0,0) -- (0,-1) -- (1,-1) -- (1,-2) -- (2,-2) -- (2,-2.5);
\node[rotate=315] at (2.5,-3) {\Huge $\cdots$};
\draw[ultra thick] (3,-3.5) -- (3,-4) -- (4,-4) -- (4,-5) -- (5,-5);
\draw[ultra thick] (0,0) -- (0.7,0.7);
\draw[ultra thick] (1,-1) -- (1.7,-0.3);
\draw[ultra thick] (2,-2) -- (2.7,-1.3);
\draw[ultra thick] (4,-4) -- (4.7,-3.3);
\draw[ultra thick] (0,-1) -- (-0.7,-1.7);
\draw[ultra thick] (1,-2) -- (0.3,-2.7);
\draw[ultra thick] (3,-4) -- (2.3,-4.7);
\draw[ultra thick] (4,-5) -- (3.3,-5.7);
\node at (-1.2,0) {\large {\bf $\mathbf a$}};
\node at (5.2,-5) {\large {\bf $\mathbf a$}};
\node at (0.85,0.85) {\large {$\mathbf 1$}};
\node at (1.85,-0.15) {\large {$\mathbf 2$}};
\node at (2.85,-1.15) {\large {$\mathbf 3$}};
\node at (4.85,-3) {\large {$\mathbf{N}$}};
\node at (-0.65,-2) {\large {$\mathbf{1}$}};
\node at (0.35,-3) {\large {$\mathbf{2}$}};
\node at (1.8,-4.5) {\large {$\mathbf{N-1}$}};
\node at (3.15,-6) {\large {$\mathbf{N}$}};
\node at (-0.5,0.25) {\large  {\bf $h_1$}};
\node at (0.5,-1.25) {\large  {\bf $h_2$}};
\node at (1.5,-1.75) {\large  {\bf $h_3$}};
\node at (3.5,-3.75) {\large  {\bf $h_N$}};
\node at (4.5,-4.75) {\large  {\bf $h_1$}};
\node at (-0.2,-0.5) {\large  {\bf $v$}};
\node at (0.8,-1.5) {\large  {\bf $v$}};
\node at (3.8,-4.5) {\large  {\bf $v$}};
\node at (0.6,0.25) {\large  {\bf $m$}};
\node at (1.6,-0.75) {\large  {\bf $m$}};
\node at (2.6,-1.75) {\large  {\bf $m$}};
\node at (4.6,-3.75) {\large  {\bf $m$}};
\node at (-0.1,-1.5) {\large  {\bf $m$}};
\node at (0.9,-2.5) {\large  {\bf $m$}};
\node at (2.9,-4.5) {\large  {\bf $m$}};
\node at (3.9,-5.5) {\large  {\bf $m$}};
\draw[ultra thick,red,<->] (1.05,0.95) -- (1.95,0.05);
\node[red,rotate=315] at (1.75,0.65) {{\large {\bf {$\widehat{a}_1$}}}};
\draw[ultra thick,red,<->] (2.05,-0.05) -- (2.95,-0.95);
\node[red,rotate=315] at (2.75,-0.35) {{\large {\bf {$\widehat{a}_2$}}}};
\draw[ultra thick,red,<->] (5.05,-3.05) -- (5.95,-3.95);
\node[red,rotate=315] at (5.75,-3.35) {{\large {\bf {$\widehat{a}_N$}}}};
\draw[ultra thick,blue,<->] (-1.45,-0.55) -- (-0.55,-1.45);
\node[blue,rotate=315] at (-1.1,-1.2) {{\large {\bf {$t_1$}}}};
\draw[ultra thick,blue,<->] (-0.45,-1.55) -- (0.45,-2.45);
\node[blue,rotate=315] at (-0.1,-2.2) {{\large {\bf {$t_2$}}}};
\draw[ultra thick,blue,<->] (0.55,-2.55) -- (1.45,-3.45);
\node[blue,rotate=315] at (0.9,-3.2) {{\large {\bf {$t_3$}}}};
\draw[ultra thick,blue,<->] (2.55,-4.55) -- (3.45,-5.45);
\node[blue,rotate=315] at (2.9,-5.2) {{\large {\bf {$t_N$}}}};
\draw[dashed] (-0.7,-1.7) -- (-1.2,-2.2);
\draw[dashed] (5,-5) -- (3.25,-6.75);
\draw[ultra thick,red,<->] (3.2,-6.7) -- (-1.1,-2.2);
\node[red] at (0.9,-4.7) {{\large {\bf {$S$}}}};
\draw[dashed] (4,-5) -- (-1.5,-5);
\draw[dashed] (-0.7,-1.7) -- (-1.5,-1.7);
\draw[ultra thick,red,<->] (-1.5,-4.95) -- (-1.5,-1.75);
\node[red,rotate=90] at (-1.75,-3.3) {{\large {\bf{$R-NS$}}}};
\draw[dashed] (1.2,0.2) -- (0.7,0.7);
\draw[ultra thick,blue,<->] (0.1,-0.9) -- (1.15,0.15);
\node[blue] at (0.8,-0.5) {{\large {\bf {$\tau$}}}};
\end{tikzpicture}}}
\caption{\sl Web diagram of $X_{N,1}$.}
\label{Fig:N1web}
\end{center}
${}$\\[-1.5cm]
\end{wrapfigure}
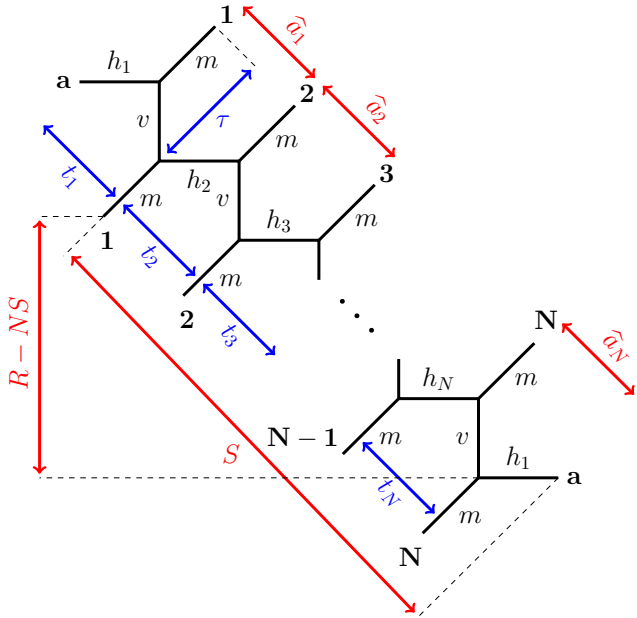

We study the topological string partition function $\mathcal{Z}_{N,1}$ (for $N\in \mathbb{N}$) of a class of toric Calabi-Yau threefolds which were called $X_{N,M=1}$ in \cite{Hohenegger:2016eqy} and which captures the partition function of a class of little string theories of A-type, as explained in the introduction. The latter are characterised through their web diagram, which is shown in \figref{Fig:N1web} for generic $N$. The boldface

\noindent
labels $\mathbf{a}$ and $\mathbf{1},\ldots,\mathbf{N}$ indicate how the external lines are mutually identified (\emph{i.e.} glued together). $X_{N,1}$ is characterised by $N+2$ K\"ahler parameters. Two choices of independent sets are shown in \figref{Fig:N1web}. The parametrisation $(t_{1,\ldots,N},m,\tau)$ (shown in blue in \figref{Fig:N1web}) has been mainly used in \cite{Hohenegger:2015btj} (among others), while the parametrisation $(\widehat{a}_{1,\ldots,N},S,R)$ (shown in red in \figref{Fig:N1web}) has been used in \cite{Bastian:2017ing,Bastian:2018jlf}. Specifically, they are related as follows
\begin{align}
(\widehat{a}_1,\ldots,\widehat{a}_N,S,R)^T=T\cdot  (t_1,\ldots,t_N,m,\tau)^T\,,\nonumber
\end{align}
where $T$ is an $(N+2)\times (N+2)$ matrix, which can be decomposed as $T=\mathcal{G}_\infty(N)\cdot S$. Here $\mathcal{G}_\infty(N)$ and $S$ are two elements in the non-perturbative symmetry group $ \widetilde{\mathbb{G}}(N)\cong\mathbb{G}(N)\times \text{Dih}_N$ found in \cite{Bastian:2018jlf}, which we shall discuss in detail in Subsection~\ref{Sect:ReviewDihedral}. Explicitly, we have for $S\in \text{Dih}_N$
\begin{align}
S=\left(\begin{array}{cc} 0 & 1\!\!1_{(N+1)\times (N+1)} \\ 1 & 0 \end{array}\right)\,.
\end{align}
while $\mathcal{G}_\infty(N)=\mathcal{G}_2(N)\cdot\mathcal{G}'_2(N) \in\mathbb{G}(N)$ where the latter matrices are given in eq.~(\ref{DefGinfGeneral}) below. Invariance of the free energy under $\widetilde{\mathbb{G}}_N$ implies that many results discussed in the literature in the basis $(t_1,\ldots,t_N,m,\tau)$ can be directly translated into the basis $(\widehat{a}_1,\ldots,\widehat{a}_N,S,R)$.


\subsection{Topological String Partition Function and Free Energy}\label{Sect:ReviewTopStringPF}
With the explicit form of the web digram given in \figref{Fig:N1web}, the (refined) topological string partition function $\mathcal{Z}_{N,1}$ associated with $X_{N,1}$ can be computed using the (refined) topological vertex formalism \cite{Aganagic:2003db, Iqbal:2007ii}. Indeed, using the parametrisation $(\widehat{a}_{1,\ldots,N},S,R)$ in \figref{Fig:N1web}, the partition function can be written in the form \cite{Bastian:2017ing}
\begin{align}
&\mathcal{Z}_{N,1}(\widehat{a}_{1,\ldots,N},S,R;\epsilon_{1,2})=\sum_{\{\alpha\}}\left(\prod_{i=1}^N Q_{m}^{|\alpha_i|}\right)\,W^{\alpha_1,\ldots,\alpha_N}_{\alpha_{1},\ldots,\alpha_{N}}(\widehat{a}_{1,\ldots,N},S;\epsilon_{1,2})\,,\label{DefPartFctGeneral}
\end{align}
where the summation is over $N$ integer partitions $\alpha_{1,\ldots,N}$ and the building blocks $W^{\alpha_1,\ldots,\alpha_N}_{\alpha_{1},\ldots,\alpha_{N}}$ are given by
\begin{align}
&W^{\alpha_1,\ldots,\alpha_N}_{\alpha_{1},\ldots,\alpha_{N}}(\widehat{a}_{1,\ldots,N},S;\epsilon_{1,2})=W_\emptyset^N(\widehat{a}_{1,\ldots,N})\,\left[\frac{\left(t/q\right)^{\frac{N-1}{2}}}{Q_\rho^{N-1}}\right]^{|\alpha_1|+\ldots+|\alpha_N|} \prod_{i,j=1}^N\frac{\vartheta_{\alpha_i,\alpha_j}(\widehat{Q}_{i,j};\rho)}{\vartheta_{\alpha_i,\alpha_j}(\bar{Q}_{i,j}\,\sqrt{q/t};\rho)}\,.\nonumber
\end{align}
Here we have used the notation (with $\rho=\sum_{k=1}^N\widehat{a}_k$)
\begin{align}
&Q_{m}=e^{2\pi im}\,,&&Q_\rho=e^{2\pi i\sum_{k=1}^N\widehat{a}_k}\,,&&q=e^{2\pi i \epsilon_1}\,,&&t=e^{2\pi i\epsilon_2}\,,&&\widehat{Q}_{i,j}=e^{-z_{ij}}\,,&&\bar{Q}_{i,j}=e^{-w_{ij}}\,,\nonumber
\end{align}
and we refer the reader to \cite{Bastian:2017ing} for a precise definition of the class of theta-functions $\vartheta_{\alpha_i,\alpha_j}$ and to \cite{Bastian:2018jlf} for a definition of $z_{ij}$ and $w_{ij}$.\footnote{We remark that our conventions for $\epsilon_{1,2}$ is different than in \cite{Haghighat:2013gba,Bastian:2017jje}.}

After the definition of the partition function, we can also introduce the free energy as the plethystic logarithm of $\mathcal{Z}_{N,1}$ (see \cite{Feng:2007ur}). Indeed, in the basis $(\widehat{a}_{1,\ldots,N},S,R)$ we have
\begin{align}
F_{N,1}(\widehat{a}_{1,\ldots,N},S,R;\epsilon_{1,2})=\text{PLog}\,\mathcal{Z}_{N,1}(\widehat{a}_{1,\ldots,N},S,R;\epsilon_{1,2})
=\sum_{k=1}^\infty\frac{\mu(k)}{k}\,\ln\mathcal{Z}_{N,1}(k\,\widehat{a}_{1,\ldots,N},k\,S,k\,R;k\,\epsilon_{1,2})\,. \label{PletFreeEnergy}
\end{align}
where $\mu$ denotes the M\"obius function. The latter can be expanded in the following fashion
\begin{align}
F_{N,1}(\widehat{a}_1,\ldots,\widehat{a}_N,S,R;\epsilon_{1,2})=\sum_{r=0}^\infty\sum_{i_1,\ldots,i_N}^\infty\sum_{k\in\mathbb{Z}}f_{i_1,\ldots,i_N,k,r}(\epsilon_1,\epsilon_2)\,\widehat{Q}_1^{i_1}\ldots \widehat{Q}_N^{i_N}\,Q_S^k\,Q_R^r\,, \label{RedSecPlog}
\end{align}
where we denoted $Q_{\widehat{a}_i}=e^{2\pi i \widehat{a}_i}$ (for $i=1,\ldots,N$), $Q_S=e^{2\pi i S}$ and $Q_R=e^{2\pi i R}$. 
For later use, we also introduce the expansion in the deformation parameters $\epsilon_{1,2}$ 
\begin{align}
F_{N,1}(\widehat{a}_1,\ldots,\widehat{a}_N,S,R;\epsilon_{1,2})=\sum_{s_1,s_2=0}^\infty\sum_{r=0}^\infty\sum_{i_1,\ldots,i_N}^\infty\sum_{k\in\mathbb{Z}}\epsilon_{1}^{s_1-1}\epsilon_{2}^{s_2-1}f^{(s_1,s_2)}_{i_1,\ldots,i_N,k,r}\,\widehat{Q}_1^{i_1}\ldots \widehat{Q}_N^{i_N}\,Q_S^k\,Q_R^r\,,\label{TaylorFreeEnergy}
\end{align}
where $f^{(s_1,s_2)}_{i_1,\ldots,i_N,k,r}\in\mathbb{Z}$ are pure constants.\footnote{Notice, since $(\widehat{a}_{1,\ldots,N},S,R)$ and $(t_{1,\ldots,N},m,\tau)$ are related through an element of the symmetry group $\widetilde{\mathbb{G}}(N)$, the expansion of the free energy in the two bases is perfectly equivalent.} In this paper, we are mostly in the coefficients with 
\begin{align}
i_1=i_2=\ldots=i_N\,.\label{Identification}
\end{align}
From the perspective of the M-brane setup which can be used to engineer the little string theories discussed in this work, these coefficients count configurations of $N$ parallel M5-branes with the same number of $i_1=i_2=\ldots=i_N$ M2-branes stretched between them respectively.\footnote{See Sections~\ref{Sect:Recursion} and \ref{Sect:ParaEnhancement} for more details on these brane configurations.} In \cite{Ahmed:2017hfr}, taking the dual point of view of monopole strings, these states were characterised by carrying the charge vector $(i_1,\ldots,i_1)$. In the companion paper \cite{Companion}, we present further results for the remaining contributions (\emph{i.e.} coefficients $f^{(s_1,s_2)}_{i_1,\ldots,i_N,k,r}$ that do not satisfy (\ref{Identification})).

From the coefficients $f^{(s_1,s_2)}_{i_1,\ldots,i_N,k,n}$ that satisfy (\ref{Identification}), we can construct two different series expansions
\begin{align}
G_{(s_1,s_2)}^{(n,N)}(R,S)&=\sum_{r=0}^\infty\sum_{k\in \mathbb{Z}}f^{(s_1,s_2)}_{\underbrace{\text{\scriptsize $n,\ldots,n$}}_{N\text{ times}},k,r}\, Q_S^kQ_R^r\,, &&\forall n\in\mathbb{N}\,,\label{DefinitionG}\\
H_{(s_1,s_2)}^{(r,N)}(\rho,S)&=\sum_{n=0}^\infty\sum_{k\in\mathbb{Z}} f^{(s_1,s_2)}_{\underbrace{\text{\scriptsize $n,\ldots,n$}}_{N\text{ times}},k,r}\,Q_S^k\,Q_\rho^n\,,&&\forall r\in\mathbb{N}\,,\label{DefinitionH}
\end{align}
with 
\begin{align}
&G_{(s_1,s_2)}^{(n,N)}(R,S)=0=H_{(s_1,s_2)}^{(r,N)}(\rho,S)\,,&&\forall s_1+s_2\in\mathbb{Z}_{\text{odd}}\,.
\end{align}
Here $Q_\rho=e^{2 \pi i \rho}$ for $\rho\in \mathbb{C}$ (which can be thought of as $\rho=\sum_{i=1}^N\widehat{a}_i$) and $Q_S=e^{2 \pi i S}$. Notice that both of these objects only capture a subset of the BPS states of the theory defined through $X_{N,1}$ and consequently also only depend on the reduced set of (K\"ahler)parameters $(R,S)$ and $(\rho,S)$ respectively. For both objects many symmetries and dualities have already been observed in \cite{Hohenegger:2015btj,Ahmed:2017hfr}. We will briefly review the known structures in the following Section~\ref{Sect:ReviewDihedral} and extend (and systematise) them in Section~\ref{Sect:ParamodularEnhancement}.

\subsection{Modular Properties and Dihedral Symmetry}\label{Sect:ReviewDihedral}
\subsubsection{$SL(2,\mathbb{Z})$ Transformations}
The two expansions (\ref{DefinitionG}) and (\ref{DefinitionH}) of (subsectors of) the free energy are quasimodular Jacobi forms under two different $SL(2,\mathbb{Z})$ symmetries that act in the following manner on the reduced set of parameters 
\begin{align}
&SL(2,\mathbb{Z})_R:&&(R,S,\rho)\longrightarrow \left(\frac{aR+b}{cR+d},\frac{S}{cR+d},\rho-\frac{c N S^2}{cR+d}\right)\,,\nonumber\\
&SL(2,\mathbb{Z})_\rho:&&(R,S,\rho)\longrightarrow \left(R-\frac{cS^2}{c\rho+d},\frac{S}{c\rho+d},\frac{a\rho+b}{c\rho+d}\right)\,,\label{SL2Actions}
\end{align}
for $a,b,c,d\in\mathbb{Z}$ with $ad-bc=1$. As explained in \cite{Hohenegger:2015cba}\footnote{The discussion of modular properties in \cite{Hohenegger:2015cba} is at the level of the partition function $\mathcal{Z}_{N,1}$, as defined in (\ref{DefPartFctGeneral}). However, similar relations also apply to the free energy $\mathcal{F}_{N,1}$.}, these modular groups also act in a non-trivial fashion on the deformation parameters $\epsilon_{1,2}$. However, the partition function $\mathcal{Z}_{N,1}$ (and similarly the free energy $\mathcal{F}_{N,1}$) fails to be a multivariable (quasi-)Jacobi form, since it does not behave properly under shifts of $\epsilon_{1,2}$ with respect to the modular parameters $\rho$ and $R$ (see eq.~(\ref{ShiftTerm}) in appendix~\ref{App:ModularStuff}). Since we are working with the Taylor expansion with respect to $\epsilon_{1,2}$ of the free energy (see (\ref{TaylorFreeEnergy})) we will not be sensitive to the details of this transformations, except for a dependence of the weight of $G^{(n,N)}_{s_1,s_2}$ and $H^{(r,N)}_{s_1,s_2}$ on $s_{1,2}$: indeed, it was observed in \cite{Hohenegger:2015btj} that $G_{(s_1,s_2)}^{(n,N)}(R,S)$ is a quasi-Jacobi form of weight $s_1+s_2-2$ and index $Nn$ under the congruence subgroup $\Gamma_0(\subg(n))\subset SL(2,\mathbb{Z})_R$, while $H_{(s_1,s_2)}^{(r,N)}(\rho,S)$ is a quasi-Jacobi form of weight $s_1+s_2-2$ and index $Nr$ under the congruence subgroup $\Gamma_0(\subg(r))\subset SL(2,\mathbb{Z})_\rho$. In both cases the function $\subg:\,\mathbb{N}\to \mathbb{N}$ was defined in (\ref{PrimeFunct}) in appendix~\ref{App:ModularStuff}.

More precisely, $G_{(s_1,s_2)}^{(n,N)}(R,S)$ can be expanded in the following fashion
\begin{align}
G_{(s_1,s_2)}^{(n,N)}(R,S)=\sum_{u=0}^{Nn}g_u^{(s_1,s_2)}(R)\,(\phi_{0,1}(R,S))^{Nn-u}\,(\phi_{-2,1}(R,S))^{u}\,,\label{ExpansionG}
\end{align}
where $g_u^{(s_1,s_2)}(R)$ is a quasi-modular form with weight $s_1+s_2+2(u-1)$ that can be written as a polynomial in the Eisenstein series $\{E_2(p_i R),E_4(p_i R),E_6(p_iR)\}$ for all the prime factors $p_i$ appearing in the decomposition of $n$. In the same fashion, $H_{(s_1,s_2)}^{(r,N)}(\rho,S)$ can be expanded in the following form
\begin{align}
H_{(s_1,s_2)}^{(r,N)}(\rho,S)=\sum_{u=0}^{Nr}h^{(s_1,s_2)}_u(\rho)\,(\phi_{0,1}(\rho,S))^{Nr-u}\,(\phi_{-2,1}(\rho,S))^{u}\,,\label{ExpansionH}
\end{align}
where $h^{(s_1,s_2)}_u(\rho)$ is a quasi-modular form of weight $s_1+s_2+2(u-1)$ that can be written as a polynomial in the Eisenstein series $\{E_2(p_i\rho),E_4(p_i,\rho),E_6(p_i\rho)\}$ for all the $p_i$ appearing in the prime factor decomposition of $r$.\footnote{We have $g_u^{(s_1,s_2)}(R)=0$ and $h^{(s_1,s_2)}_u(\rho)=0$ if $u<0$.} Notice, due to the presence of $E_2$ in (\ref{ExpansionG}) and (\ref{ExpansionH}), in general neither $G_{(s_1,s_2)}^{(n,N)}(R,S)$ nor $H_{(s_1,s_2)}^{(r,N)}(\rho,S)$ are strictly speaking Jacobi forms, but are rather quasi-Jacobi forms.

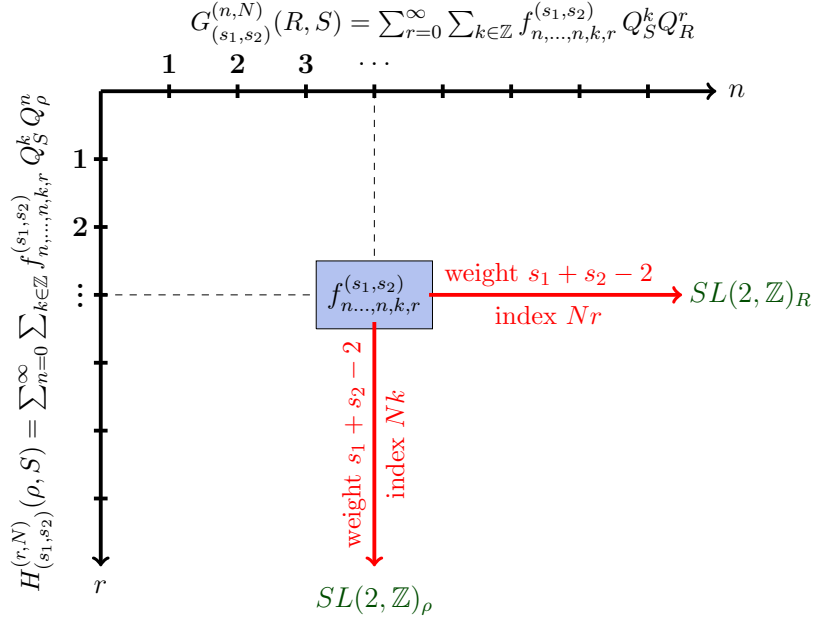
\begin{figure}[h]
\begin{center}
\scalebox{0.9}{\parbox{12cm}{\begin{tikzpicture}[scale = 1]
\node at (3,1) {\small $G_{(s_1,s_2)}^{(n,N)}(R,S)=\sum_{r=0}^\infty\sum_{k\in \mathbb{Z}}f^{(s_1,s_2)}_{n,\ldots,n,k,r}\, Q_S^kQ_R^r$};
\node[rotate=90] at (-3,-3.7) {\small $H_{(s_1,s_2)}^{(r,N)}(\rho,S)=\sum_{n=0}^\infty\sum_{k\in\mathbb{Z}} f^{(s_1,s_2)}_{n,\ldots,n,k,r}\,Q_S^k\,Q_\rho^n$};
\draw[ultra thick,->] (-2,0) -- (7,0);
\draw[ultra thick] (-1,0.1) -- (-1,-0.1);
\node at (-1,0.4) {\bf 1};
\draw[ultra thick] (0,0.1) -- (0,-0.1);
\node at (0,0.4) {\bf 2};
\draw[ultra thick] (1,0.1) -- (1,-0.1);
\node at (1,0.4) {\bf 3};
\draw[ultra thick] (2,0.1) -- (2,-0.1);
\node at (2,0.4) {\bf $\ldots$};
\draw[ultra thick] (3,0.1) -- (3,-0.1);
\draw[ultra thick] (4,0.1) -- (4,-0.1);
\draw[ultra thick] (5,0.1) -- (5,-0.1);
\draw[ultra thick] (6,0.1) -- (6,-0.1);
\node at (7.3,0) {$n$};
\draw[ultra thick,->] (-2,0) -- (-2,-7);
\draw[ultra thick] (-2.1,-1) -- (-1.9,-1);
\node at (-2.3,-1) {\bf 1};
\draw[ultra thick] (-2.1,-2) -- (-1.9,-2);
\node at (-2.3,-2) {\bf 2};
\draw[ultra thick] (-2.1,-3) -- (-1.9,-3);
\node at (-2.3,-2.9) {\bf $\vdots$};
\draw[ultra thick] (-2.1,-4) -- (-1.9,-4);
\draw[ultra thick] (-2.1,-5) -- (-1.9,-5);
\draw[ultra thick] (-2.1,-6) -- (-1.9,-6);
\node at (-2,-7.3) {$r$};
\draw[dashed] (2,0) -- (2,-3);
\draw[dashed] (-2,-3) -- (2,-3);
\draw[fill=green!20!blue!30!white, xshift=1.5cm, yshift=-3.5cm] (-0.35,0) -- (1.35,0) -- (1.35,1) -- (-0.35,1) -- (-0.35,0);
\node[xshift=1.5cm, yshift=-3.5cm] at (0.5,0.5) {$f_{n\ldots,n,k,r}^{(s_1,s_2)}$}; 
\draw[ultra thick, red,->] (2.8,-3) -- (6.5,-3);
\node[red] at (4.55,-2.7) {\small weight $s_1+s_2-2$};
\node[red] at (4.55,-3.3) {\small index $Nr$};
\node[green!30!black] at (7.5,-3) {$SL(2,\mathbb{Z})_R$};
\draw[ultra thick, red,->] (2,-3.4) -- (2,-7);
\node[rotate=90,red] at (1.7,-5.2) {\small weight $s_1+s_2-2$};
\node[rotate=90,red] at (2.3,-5.2) {\small index $Nk$};
\node[green!30!black] at (2,-7.5) {$SL(2,\mathbb{Z})_\rho$};
\end{tikzpicture}}}
\end{center} 
\caption{{\it Graphical overview of modular transformations for quasi-Jacobi forms created from the coefficients $f_{n,\ldots,n,k,r}^{(s_1,s_2)}$.}}
\label{fig:OverviewQuasiJacobi}
\end{figure}

\subsubsection{Dihedral Transformations}\label{Sect:DihedralSymmetry}
It is important to stress that both $G_{(s_1,s_2)}^{(n,N)}(R,S)$ and $H_{(s_1,s_2)}^{(r,N)}(\rho,S)$ are composed from the same coefficients $f_{n,\ldots,n,k,r}^{(s_1,s_2)}$, however, resummed in two different fashions. In order to help keeping track of the various properties, we provide a graphical representation in \figref{fig:OverviewQuasiJacobi}. As was argued in \cite{Bastian:2018jlf}, besides the modular properties of $G_{(s_1,s_2)}^{(n,N)}(R,S)$, the coefficients $f^{(s_1,s_2)}_{i_1,\ldots,i_N,k,r}$ also satisfy additional identities of the form 
\begin{align}
&f^{(s_1,s_2)}_{i_1,\ldots,i_N,k,r}=f^{(s_1,s_2)}_{i'_1,\ldots,i'_N,k',r'}&&\text{for}&&\left\{\begin{array}{l}(i'_1,\ldots,i'_N,k',r')^T=G^T\cdot (i_1,\ldots,i_N,k,r)^T\,, \\ s_1,s_2\in\mathbb{N}\cup \{0\}\,.\end{array}\right.\label{DihedralAction}
\end{align}
Here $G$ are $(N+2)\times (N+2)$ matrices that form the group $\widetilde{G}(N)\cong \mathbb{G}(N)\times \text{Dih}_N$ with
\begin{align}
\mathbb{G}(N)\cong\left\{\begin{array}{lcl}\text{Dih}_3 & \text{if} & N=1\,, \\ \text{Dih}_N & \text{if} & N=2,3\,,\\\text{Dih}_\infty & \text{if} & N\geq 4\,.\end{array}\right.
\end{align}
For any $N$, this group can be understood as being generated by two elements of order 2
\begin{align}
\mathbb{G}(N)\cong\left\langle\{\mathcal{G}_2(N),\mathcal{G}'_2(N)\big|(\mathcal{G}(N))^2=(\mathcal{G}'(N))^2=(\mathcal{G}(N)\cdot\mathcal{G}'(N))^n=1\!\!1_{(N+2)\times (N+2)}\}\right\rangle\,,
\end{align} 
where $n=3$ for $N=1$, $n=N$ for $N=2,3$ and $n\to \infty$ for $N\geq 4$. The generating group elements can explicitly be written as
\begin{align}
&\mathcal{G}_2(N)=\left(\begin{array}{ccccc} & & & 0 & 0 \\ & 1\!\!1_{N\times N} & & \vdots & \vdots \\ & & & 0 & 0 \\ 1 & \cdots & 1 & -1 & 0 \\ N & \cdots & N & -2N & 1 \end{array}\right)\,,&&\text{and}&&\mathcal{G}'_2(N)=\left(\begin{array}{ccccc} & & & -2 & 1 \\ & 1\!\!1_{N\times N} & & \vdots & \vdots \\ & & & -2 & 1 \\ 0 & \cdots & 0 & -1 & 1 \\ 0 & \cdots & 0 & 0 & 1 \end{array}\right)\,.\label{DefGinfGeneral}
\end{align}

\subsection{Recursion and Torus Orbifold}\label{Sect:Recursion}
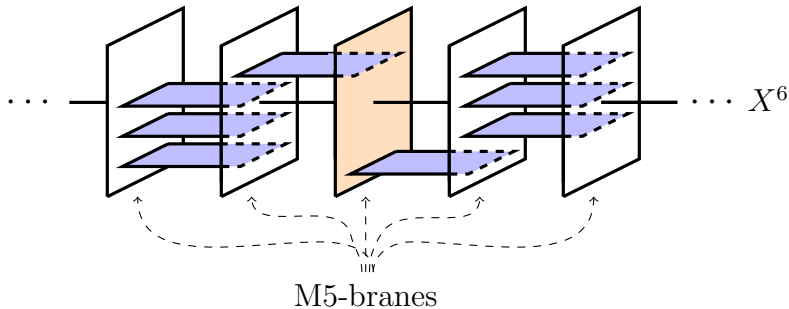
\begin{figure}[h]
\begin{center}
\scalebox{1}{\parbox{10.6cm}{\begin{tikzpicture}[scale = 1]
\draw[very thick] (0,0) -- (1,0.5) -- (1,2.5) -- (0,2) -- (0,0);
\draw[very thick,xshift=1.5cm] (0,0) -- (1,0.5) -- (1,2.5) -- (0,2) -- (0,0);
\draw[yshift=-0.6cm,fill=blue!25!white,very thick] (0.8,1.3) -- (0.2,1) -- (1.5,1) -- (1.5,1.3) -- (0.8,1.3);
\draw[yshift=-0.6cm,fill=blue!25!white,very thick,dashed] (1.5,1) --(1.75,1) -- (2.35,1.3) -- (1.5,1.3);
\draw[yshift=-0.2cm,fill=blue!25!white,very thick] (0.8,1.3) -- (0.2,1) -- (1.5,1) -- (1.5,1.3) -- (0.8,1.3);
\draw[yshift=-0.2cm,fill=blue!25!white,very thick,dashed] (1.5,1) --(1.75,1) -- (2.35,1.3) -- (1.5,1.3);
\draw[yshift=0.2cm,fill=blue!25!white,very thick] (0.8,1.3) -- (0.2,1) -- (1.5,1) -- (1.5,1.3) -- (0.8,1.3);
\draw[yshift=0.2cm,fill=blue!25!white,very thick,dashed] (1.5,1) --(1.75,1) -- (2.35,1.3) -- (1.5,1.3);
\draw[very thick] (1.5,0.2) -- (1.5,2);
\draw[yshift=0.6cm,fill=blue!25!white,very thick,xshift=1.5cm] (0.8,1.3) -- (0.2,1) -- (1.5,1) -- (1.5,1.3) -- (0.8,1.3);
\draw[very thick,xshift=3cm,fill=orange!25!white] (0,0) -- (1,0.5) -- (1,2.5) -- (0,2) -- (0,0);
\draw[yshift=0.6cm,fill=blue!25!white,very thick,dashed,,xshift=1.5cm] (1.5,1) --(1.75,1) -- (2.35,1.3) -- (1.4,1.3);
\draw[very thick,xshift=1.5cm] (1.5,0.5) -- (1.5,2);
\draw[very thick,xshift=4.5cm] (0,0) -- (1,0.5) -- (1,2.5) -- (0,2) -- (0,0);
\draw[yshift=-0.7cm,fill=blue!25!white,very thick,xshift=3cm] (0.8,1.3) -- (0.2,1) -- (1.5,1) -- (1.5,1.3) -- (0.8,1.3);
\draw[yshift=-0.7cm,fill=blue!25!white,very thick,dashed,xshift=3cm] (1.5,1) --(1.75,1) -- (2.35,1.3) -- (1.5,1.3);
\draw[very thick,xshift=6cm] (0,0) -- (1,0.5) -- (1,2.5) -- (0,2) -- (0,0);
\draw[yshift=0.6cm,fill=blue!25!white,very thick,xshift=4.5cm] (0.8,1.3) -- (0.2,1) -- (1.5,1) -- (1.5,1.3) -- (0.8,1.3);
\draw[yshift=0.6cm,fill=blue!25!white,very thick,dashed,xshift=4.5cm] (1.5,1) --(1.75,1) -- (2.35,1.3) -- (1.5,1.3);
\draw[yshift=0.2cm,fill=blue!25!white,very thick,xshift=4.5cm] (0.8,1.3) -- (0.2,1) -- (1.5,1) -- (1.5,1.3) -- (0.8,1.3);
\draw[yshift=0.2cm,fill=blue!25!white,very thick,dashed,xshift=4.5cm] (1.5,1) --(1.75,1) -- (2.35,1.3) -- (1.5,1.3);
\draw[yshift=-0.2cm,fill=blue!25!white,very thick,xshift=4.5cm] (0.8,1.3) -- (0.2,1) -- (1.5,1) -- (1.5,1.3) -- (0.8,1.3);
\draw[yshift=-0.2cm,fill=blue!25!white,very thick,dashed,xshift=4.5cm] (1.5,1) --(1.75,1) -- (2.35,1.3) -- (1.5,1.3);
\node at (-1,1.25) {\large $\cdots$}; 
\draw[very thick] (-0.5,1.25) -- (0,1.25);
\draw[very thick,xshift=1.5cm] (0.5,1.25) -- (1.5,1.25);
\draw[very thick,xshift=3cm] (0.5,1.25) -- (1.5,1.25);
\draw[very thick,xshift=6cm] (0.5,1.25) -- (1.5,1.25);
\draw[very thick,xshift=1.5cm] (5,1.25) -- (6,1.25);
\node at (3.4,-1.3) {\text{M5-branes}};
\draw[dashed,->] (3.4,-1) -- (3.4,0);
\draw[dashed,->] (3.45,-1) to [out=90,in=240] (3.6,-0.5) to [out=60,in=270] (4.9,0);
\draw[dashed,->] (3.5,-1) to [out=90,in=180] (4,-0.6) to [out=0,in=270] (6.4,0);
\draw[dashed,->] (3.35,-1) to [out=90,in=285] (3.2,-0.5) to [out=105,in=270] (1.9,0);
\draw[dashed,->] (3.5,-1) to [out=90,in=0] (2.8,-0.6) to [out=180,in=270] (0.4,0);
\node at (8,1.25) {\large $\cdots$}; 
\node at (8.7,1.3) {$X^6$};
\draw[very thick,xshift=3cm] (1.5,0.2) -- (1.5,2);
\draw[very thick,xshift=4.5cm] (1.5,0.2) -- (1.5,2);
\end{tikzpicture}}}
\end{center} 
\caption{{\it Configuration of 5 M5-branes with various M2-branes (shown in blue) stretched between them. The M5-brane in the middle (shown in orange) has only one M2-brane ending on either side. The labelling of the coordinate $X^6$ is explained later on in Section~\ref{Sect:ParaEnhancement}.}}
\label{fig:PartM5braneConfig}
\end{figure}

In \cite{Hohenegger:2015btj} and \cite{Ahmed:2017hfr} respectively, two further properties of the free energy were observed. In \cite{Hohenegger:2015btj}, by studying numerous explicit examples, it was conjectured that the following relation holds for generic $N>1$
\begin{align}
\sum_{s_1,s_2=0}^\infty \epsilon_1^{2s_1-1}\epsilon_2^{2s_2-1}\,G_{(s_1,s_2)}^{(1,N)}(R,S)=&\,W(R,S,\epsilon_{1,2})\sum_{s_1,s_2=0}^\infty \epsilon_1^{s_1-1}\epsilon_2^{s_2-1}\,G_{(s_1,s_2)}^{(1,N-1)}(R,S)\nonumber\\
&+W(R,S,\epsilon_{1,2})^{N-1}\sum_{s_1,s_2=0}^\infty \epsilon_1^{s_1-1}\epsilon_2^{s_2-1}K_{(s_1,s_2)}(R,S)\,.\label{ReductionEqW}
\end{align}
Here $K_{(s_1,s_2)}(R,S)$ corresponds to a contribution to the non-compact free energy (which counts BPS states corresponding to a single M2-brane stretched between two M5-branes), which in our notation can be written as
\begin{align}
K_{(s_1,s_2)}(R,S)&=\sum_{r=0}^\infty\sum_{k\in \mathbb{Z}}f^{(s_1,s_2)}_{1,0,k,r}\, Q_S^kQ_R^r\,,
\end{align}
and is given explicitly by
\begin{align}
\sum_{s_1,s_2=0}^\infty \epsilon_1^{s_1-1}\epsilon_2^{s_2-1}K_{(s_1,s_2)}(R,S)=\frac{\theta_1(R;S+\epsilon_+)\theta_1(R;S-\epsilon_+)}{\theta_1(R;\epsilon_1)\theta_1(R;\epsilon_2)}\,.
\end{align}
Here $\theta_1$ are Jacobi theta functions and $\epsilon_\pm=\frac{\epsilon_1\pm \epsilon_2}{2}$. Finally, an explicit form for the function $W$ is given by
\begin{align}
W(R,S,\epsilon_{1,2})=\frac{\theta_1(R;S+\epsilon_-)\theta_1(R;S-\epsilon_-)-\theta_1(R;S+\epsilon_+)\theta_1(R;S-\epsilon_+)}{\theta_1(R;\epsilon_1)\theta_1(R;\epsilon_2)}\,.\label{FormExplicitW}
\end{align} 
The latter function has also appeared in the context of counting of BPS states corresponding to configurations of parallel M5-branes along a non-compact direction with M2-branes stretched between them \cite{Hohenegger:2015cba}: in the NS-limit ($\epsilon_2\to 0$), the BPS counting function for a configuration with an M5-brane that has only one M2-brane ending on it on either side (see \figref{fig:PartM5braneConfig} for an example) is up to a factor of $\lim_{\epsilon_2\to 0 } W(R,S,\epsilon_{1,2})$ proportional to the BPS counting function of the same configuration, where this brane has been removed. This result is similar in nature to (\ref{ReductionEqW}). Notice in particular that $F_{(s_1,0)}(R,S)=G^{(1,1)}_{(s_1,0)}(R,S)$, which is pertinent for the NS-limit. In this limit, (\ref{ReductionEqW}) encodes various relations among different coefficients $f_{i_1,\ldots,i_N,k,r}^{(s_1,s_2)}$ which still have to be compatible with the modular properties and other symmetries as discussed in Section~\ref{Sect:ReviewDihedral}. As we shall discuss in Section~\ref{Sect:Reconstruction}, this (as well as other symmetries) poses strong constraints on the function $W(R,S,\epsilon_{1,2})$, and in fact forces the first few terms in an expansion in $\epsilon_{1,2}$ to essentially take the form (\ref{FormExplicitW}).

A further observation was made in \cite{Ahmed:2017hfr}, where in many examples it was shown that the reduced free energy exhibits a particular Hecke structure. In the notation used in the present work, the latter can be formulated as follows (with $G_{(s_1,0)}^{(n,N)}(R,S)$ defined in (\ref{DefinitionG}) 
\begin{align}
&G_{(s_1,0)}^{(n,N)}(R,S)=\mathcal{T}_n\, \big( G_{(s_1,0)}^{(1,N)}(R,S) \big)\,,&& \forall n,N\in\mathbb{N}\,.\label{NSLimitHeckeRelation}
\end{align}
Here $\mathcal{T} _n$ is an operator which can be expressed as a sum over Hecke operators: let $f_{w,\vec{r}}(R,S)$ be a multivariable Jacobi form of weight $w$ and index vector $\vec{r}$, then
\begin{align}
\mathcal{T} _n(f_{w,\vec{r}})=\sum_{k|n}k^{w-1}\,\mu(k)\,\mathcal{H}_{\frac{n}{k}}(f_{w,\vec{r}}(kR,kS))\,,\label{HeckeGeneration}
\end{align}
where the $\mu$ denotes the M\"obius function as before and $\mathcal{H}_{k}$ the $k$th Hecke operator which maps $f_{w,\vec{r}}$ to a Jacobi form of weight $w$ and index vector $k\vec{r}$ (see (\ref{DefHeckeGeneric}) for the explicit definition). The relation (\ref{HeckeGeneration}) together with (\ref{ReductionEqW}) essentially fixes the $G_{(s_1,0)}^{(n,N)}(R,S)$ in the NS-limit. Furthermore, it was argued in \cite{Ahmed:2017hfr} that a structure such as (\ref{HeckeGeneration}) is the hallmark of the free energy of a symmetric orbifold CFT~\cite{Dijkgraaf:1996xw}. It was thus conjectured that the BPS states contributing to the $G_{(s_1,0)}^{(n,N)}(R,S)$ form a symmetric orbifold sigma model, whose target space is the $N$th symmetric product of moduli spaces of monopole strings of charge $(1,\ldots,1)$. In Section~\ref{Sect:RelFourier} we shall discuss the conditions that (\ref{HeckeGeneration}) imposes on the coefficients $f^{(s_1,s_2)}_{i_1,\ldots,i_N,k,n}$ that appear in the expansion of $G_{(s_1,s_2)}^{(n,N)}(R,S)$ and show that they are perfectly compatible with the dihedral symmetry (and the modular properties) discussed in Section~\ref{Sect:ReviewDihedral}.
\section{Paramodular Group and Symmetry Enhancement}\label{Sect:ParamodularEnhancement}
\subsection{Paramodular Action}
In the following we shall argue that the reduced free energy 
\begin{align}
\mathcal{F}_{(s_1,s_2)}^{N}(R,S,\rho)=\sum_{r,n=0}^\infty \sum_{k\in\mathbb{Z}}f_{n,\ldots,n,k,r}^{(s_1,s_2)}\,Q_S^k Q_R^r Q_\rho^n\,,\label{DefFreeEnergyReduced}
\end{align}
supports an action of the paramodular group $\Sigma_N$ of level $N$ (see app.~\ref{App:Paramod} for the definition (\ref{DefParaModN}) and more properties of $\Sigma_N$). More precisely, we propose that $\Sigma_N$ acts on the period matrix
\begin{align}
&\Omega_N=\left(\begin{array}{cc} R & S \\ S & \rho/N\end{array}\right)\,,&&\forall N\in \mathbb{N}\,,\label{DefPeriodMatrix}
\end{align}
in the canonical manner (see (\ref{ParaGroupAction})).\footnote{By rescaling $\rho$ in (\ref{DefPeriodMatrix}) we have chosen conventions following the works \cite{Ahmed:2017hfr} and \cite{Hohenegger:2016yuv}. In this manner (\ref{DefFreeEnergyReduced}) is an integer series expansion in $Q_\rho$ (rather than $Q_\rho^N$).} To argue for this action, we simply have to check covariance of $\mathcal{F}_{(s_1,s_2)}^{N}$ under all the generators $\{J_M,S_{1,2,3}\}$. According to (\ref{ActionParaPeriod}) the action of $S_{1,2,3}$ corresponds to a shift
\begin{align}
&S_1:\,R\longmapsto R+1\,,&&S_2:\,S\longmapsto S+1\,,&&S_3:\,\rho\longmapsto \rho+1\,,
\end{align}
which indeed leaves $\mathcal{F}_{(s_1,s_2)}^{N}$ invariant. This is apparent from the Fourier expansion (\ref{DefFreeEnergyReduced}) in powers of $Q_R$, $Q_S$ and $Q_\rho$ respectively. The action of $J_N$ is more complicated and can be better understood by considering $J_N=S_R\cdot S_\rho$, where the action of the latter is given in (\ref{ActSl2s}). In this way, $S_R$ and $S_\rho$ correspond to generators of $SL(2,\mathbb{Z})_R$ and $SL(2,\mathbb{Z})_\rho$ in (\ref{SL2Actions}) respectively. In this way, covariance of $\mathcal{F}_{(s_1,s_2)}^{N}$ follows from the modular properties of $G_{(s_1,s_2)}^{(n,N)}(R,S)$ and $H_{(s_1,s_2)}^{(r,N)}(\rho,S)$ as discussed in Section~\ref{Sect:ReviewDihedral}.

A interesting question is to what extent the dihedral symmetry $\mathbb{G}(N)$ discussed in Section~\ref{Sect:DihedralSymmetry} is part of the paramodular action, or whether it in fact enhances the symmetry group. To understand this, we have to translate the dihedral action (\ref{DihedralAction}) on the Fourier coefficients of the reduced free energy, into an action on the reduced parameters $(R,S,\rho)$. To this end, we first translate the action of $\mathbb{G}(N)$ on the basis $(\widehat{a}_1,\ldots,\widehat{a}_N,S,R)$ (which is given by the matrices $\mathcal{G}_2(N)$ and $\mathcal{G}'_2(N)$ in (\ref{DefGinfGeneral})) into an action on the basis $(\rho,\widehat{a}_1,\ldots,\widehat{a}_{N-1},S,R)$, where $\rho=\sum_{i=1}^N\widehat{a}_i$. The latter is given by the matrices
\begin{align}
&\mathcal{G}_2(N)\,,&&\text{and}&\mathcal{G}_2(N)\cdot \mathcal{G}'_2(N)\cdot \mathcal{G}_2(N)\,.
\end{align}
From these matrices, we can read off the action of the generators of $\mathbb{G}(N)$ on the period matrix $\Omega_N$ (in the sense of (\ref{ParaGroupAction})), which take the form
\begin{align}
&\mathfrak{g}_2(N)=\left(\begin{array}{cccc} -1 & N & 0 & 0 \\ 0 & 1 & 0 & 0 \\ 0 & 0 & -1 & 0 \\ 0 & 0 & N & 1 \end{array}\right)\,,&&\mathfrak{g}'_2(N)=\left(\begin{array}{cccc} N-1 & -N(N-2) & 0 & 0 \\ 1 & -(N-1) & 0 & 0 \\ 0 & 0 & N-1 & 1 \\ 0 & 0 & -N(N-2) & -(N-1) \end{array}\right)\,,\nonumber
\end{align}
with $\text{det}(\mathfrak{g}_2(N))=1=\text{det}(\mathfrak{g}'_2(N))$. Following the presentation (\ref{AltPresSigmaN}) of $\Sigma_N$, the paramodular group of level $N$, we find that $\mathfrak{g}_2(N),\mathfrak{g}'_2(N)\in\Sigma_N$, such that $\mathbb{G}(N)$ does not further enhance the symmetry group of the reduced free energy.\footnote{Notice, also $\widetilde{\mathbb{G}}(N)=\mathbb{G}(N)\times \text{Dih}_N$ leaves the reduced free energy invariant, as $\text{Dih}_N$ acts trivially on $\Omega_N$.} In other words, the BPS sector contributing to the reduced free energy is invariant under $\mathbb{G}(N)\subset \Sigma_N$.


\subsection{Paramodular Group and Enhancement in the NS-Limit}\label{Sect:ParaEnhancement}
The discussion in the previous Subsection has been valid for generic $s_{1,2}\geq 0$. It has been observed in various works, that in the case $s_2=0$ (or equivalently $s_1=0$), which corresponds to the NS-limit, the reduced free energy $\mathcal{F}_{(s_1,s_2=0)}^{N}(R,S,\rho)$ acquires additional symmetries (\emph{e.g.} eq.~(\ref{NSLimitHeckeRelation}) reviewed above). Furthermore, in \cite{Hohenegger:2015cba} it was argued that in the limit $\epsilon_2\to 0$, the system of M5- and M2-branes (a subsector of) whose BPS states is counted by the reduced free energy $\mathcal{F}_{(s_1,s_2)}^{N}$ allows for an enhanced symmetry (which is part of the U-duality group). To shed further light on this, it was proposed in \cite{Hohenegger:2015btj} to further compactify the M-brane setup on $S_5^1$, \emph{i.e.} to consider M5- and M2-branes on $S^1_0\times S^1_\tau\times \mathbb{R}_{||}^3\times S^1_5\times S^1_\rho\times\mathbb{R}^4_{\perp}$
\begin{center}
\begin{tabular}{c|c|c|c|c|c|c|c|c|c|c|c}
& $S_0^1$ & $S_\tau^1$ & \multicolumn{3}{c|}{$\mathbb{R}_{||}^3$} & $S^1_5$ & $S^1_\rho$ &  \multicolumn{4}{c}{$\mathbb{R}_{\perp}^4$}  \\[4pt]\hline\hline
&&&&&&&&&&&\\[-14pt]
& $X^0$ & $X^1$ & $X^2$ & $X^3$ & $X^4$ & $X^5$ & $X^6$ & $X^7$ & $X^8$ & $X^9$ & $X^{10}$\\[4pt]\hline\hline
&&&&&&&&&&&\\[-14pt]
\text{M5-branes} & $\bullet$ & $\bullet$ & $\bullet$ & $\bullet$ & $\bullet$ & $\bullet$ & & & & & \\[4pt]\hline
&&&&&&&&&&&\\[-14pt]
\text{M2-branes} & $\bullet$ & $\bullet$ & & & & & $\bullet$ & & & & \\[4pt]\hline\hline
&&&&&&&&&&&\\[-14pt]
$\epsilon_1$\text{-deformation} & $\circ$ &  & $\circ$ &$\circ$ & & & & $\circ$ & $\circ$ & $\circ$ &  $\circ$\\[4pt]\hline
&&&&&&&&&&&\\[-14pt]
$\epsilon_2$\text{-deformation} & $\circ$ &  & & & $\circ$ &$\circ$ & & $\circ$ & $\circ$ & $\circ$ &  $\circ$\\[4pt]
\end{tabular}
\end{center}
Here we have introduced 11-dimensional coordinates $X^{0,\ldots,10}$. Besides the configuration of the branes, we have also indicated how the $\epsilon$-parameters appear as $U(1)$-deformations in this picture. Indeed, upon introducing complex coordinates $(z^1,z^2)=(X^2+iX^3,X^4+iX^5)$ and $(w_1,w_2)=(X^7+iX^8,X^9+iX^{10})$, they arise from the following $U(1)$ action with respect to the $X^0$-direction \cite{Haghighat:2013gba}
\begin{align}
&U(1)_{\epsilon_1}\times U(1)_{\epsilon_2}:&&(z^1,z^2)\longmapsto (e^{2\pi i\epsilon_1}\, z^1,e^{2\pi i \epsilon_2}\, z^2)\nonumber\\
& &&(w^1,w^2)\longmapsto (e^{-i\pi (\epsilon_1+\epsilon_2)}\, w^1,e^{-i\pi (\epsilon_1+\epsilon_2)}\, w^2)\,.
\end{align}
As was argued in \cite{Hohenegger:2015cba} for $\epsilon_1\neq0\neq \epsilon_2$ the directions $(X^1,X^6)$ are not twisted and thus form a torus $S^1_\tau\times S^1_\rho\sim T^2$ which affords an $SL(2,\mathbb{Z})$ symmetry. The latter in particular contains the generator responsible for T-duality of the little string theories \cite{Hohenegger:2015cba,Hohenegger:2016eqy,Kim:2015gha} engineered from this M-brane configuration, geometrically realised through the exchange of $S^1_\tau$ and $S^1_\rho$.\footnote{From the perspective of the web diagram, such as \figref{Fig:N1web}, this T-duality is simply realised as a rotation by 90 degrees.} Moreover, in the limit $\epsilon_2\to 0$, also the direction $X^5$ becomes untwisted thus leading to a larger U-duality group (geometrically stemming from $S^1_\tau\times S^1_5\times S^1_\rho$).\footnote{In the language of \cite{Ahmed:2017hfr} this additional symmetry acts by exchanging the gauge theory instantons with charged monopole strings.} By calculating explicit expansions of $H_{(s_1,s_2)}^{(r,N)}(\rho,S)$, in appendix~\ref{App:SeriesExpansionsOrbifoldSector}, we provide evidence that this symmetry acts on the reduced free energy through the exchange $R\longleftrightarrow \rho$, which implies at the level of the Fourier coefficients 
\begin{align}
&f^{(s_1,0)}_{n,\ldots,n,k,r}=f^{(s_1,0)}_{r,\ldots,r,k,n}\,,&&\forall n,r\geq 1\,,&&k\in\mathbb{Z}\,,&&s_1\geq 0\,.\label{RelationFourierCoefficientsFreeEnergy}
\end{align}
At the level of the paramodular group, this symmetry is realised by the generator $V_N$, which is defined in (\ref{DefExtensionParamodular}) (and its action on $\Omega_N$ is given in (\ref{ActionExtensionParamodular}). The generator $V_N$ extends $\Sigma_N$ to a subgroup of $Sp(4,\mathbb{R})$, which we denote by $\Sigma^*_N$. As was discussed in \cite{Belin:2018oza},  the partition functions of symmetric orbifold CFTs have $\Sigma^*_N$ as symmetry group. This agrees with the conclusion of \cite{Ahmed:2017hfr}
as we have the following relation,
\begin{align}
\mathcal{F}^N_{(s_1,0)}(R,S,\rho)=\sum_{n=1}^{\infty} Q_{\rho}^n \mathcal{T}_n\big( G^{(1,N)}_{(s_1,0)}(R,S) \big) &= \sum_{k=1}^{\infty} \frac{\mu(k)}{k} \Big(  \sum_{n=1}^{\infty} Q_{\rho}^{kn} \mathcal{H}_n\big(G^{(1,N)}_{(s_1,0)}(kR,kS) \big) \Big)\,.\label{THrelation}
\end{align}
Here the infinite sum over Hecke operators acting on $G^{(1,N)}_{(s_1,0)}$ in the last expression is directly related to the (full\footnote{\emph{I.e.} not only the part counting irreducible single particle states.}) free energy studied \cite{Ahmed:2017hfr}. According to \cite{Gritsenko1995}, it follows that the latter transforms suitably under the extended paramodular group. From the relation (\ref{THrelation}), we can thus see that this transformation behavior carries over to the reduced sector of ({\ref{RedSecPlog}}) as defined by the condition (\ref{Identification}).
\subsubsection{Relation Among Fourier Coefficients}\label{Sect:RelFourier}
The relation (\ref{RelationFourierCoefficientsFreeEnergy}) for the Fourier coefficients $f^{(s_1,0)}_{n,\ldots,n,k,r}$ can in fact be combined with the action of the modular groups $SL(2,\mathbb{Z})_R$ and $SL(2,\mathbb{Z})_\rho$. By studying the examples given in appendix~\ref{App:SeriesExpansionsOrbifoldSector}, we find
\begin{align}
&f^{(s_1,0)}_{\underbrace{\text{\scriptsize $n,\ldots,n$}}_{N\text{ times}},k,r}=f^{(s_1,0)}_{\underbrace{\text{\scriptsize $n',\ldots,n'$}}_{N\text{ times}},k',r'}&&\text{for} &&\left\{\begin{array}{llc}k^2-4Nnr=k'\,^2-4Nn'r' & &\\\hspace{0.5cm} \text{  and  } k=\pm k'\text{ mod }2N  & \text{if} & N\notin \mathbb{N}_{\text{prime}}\,, \\[12pt] k^2-4Nnr=k'\,^2-4Nn'r' & \text{if} & N\in \mathbb{N}_{\text{prime}}\,.\end{array}\right.\label{RelationsGenFourier}
\end{align} 
This relation among Forier coefficients is in fact also implied by (\ref{NSLimitHeckeRelation}). To see this, consider the following Fourier expansion
\begin{align}
&G_{(s_1,0)}^{(1,N)}(R,S)=\sum_{r=0}^\infty\sum_{k\in\mathbb{Z}}f^{(s_1,0)}_{\underbrace{\text{\scriptsize $1,\ldots,1$}}_{N\text{ times}},k,r}\,Q_R^r\,Q_S^k\,.
\end{align}
Since $G_{(s_1,0)}^{(1,N)}$ is a quasi-Jacobi form of index $N$ (and even weight), the Fourier coefficients satisfy
\begin{align}
&f^{(s_1,0)}_{\underbrace{\text{\scriptsize $1,\ldots,1$}}_{N\text{ times}},k,r}=f^{(s_1,0)}_{\underbrace{\text{\scriptsize $1,\ldots,1$}}_{N\text{ times}},k',r'}&&\text{for} &&\left\{\begin{array}{llc}k^2-4Nr=k'\,^2-4Nr'\text{  and  } k=\pm k'\text{ mod }2N & \text{if} & N\notin \mathbb{N}_{\text{prime}}\,, \\ k^2-4Nr=k'\,^2-4Nr' & \text{if} & N\in \mathbb{N}_{\text{prime}}\,.\end{array}\right.\label{RelationsInitialFourier}
\end{align} 
Furthermore, as observed in \cite{Ahmed:2017hfr}, the relation (\ref{NSLimitHeckeRelation}) implies the following Fourier expansion
\begin{align}
G_{(s_1,0)}^{(n,N)}(R,m)=\sum_{r=0}^\infty\sum_{k\in\mathbb{Z}}f^{(s_1,0)}_{\underbrace{\text{\scriptsize $n,\ldots,n$}}_{N\text{ times}},k,r}\,Q_R^r\,Q_m^k=\sum_{r=0}^\infty\sum_{k\in\mathbb{Z}}f^{(s_1,0)}_{\underbrace{\text{\scriptsize $1,\ldots,1$}}_{N\text{ times}},k,nr}\,Q_R^r\,Q_m^k\,.
\end{align}
Eq.~(\ref{RelationsInitialFourier}) then implies\footnote{Notice that $f^{(s_1,0)}_{n,\ldots,n,k,r}=f^{(s_1,0)}_{n,\ldots,n,-k,r}$.} eq.~(\ref{RelationsGenFourier}). Furthermore, one can check, that the action (\ref{DihedralAction}) for $G\in\{\mathcal{G}_2(N),\mathcal{G}'_2(N)\}$ in (\ref{DefGinfGeneral}) is compatible with (\ref{RelationsInitialFourier}):
{\allowdisplaybreaks
\begin{align}
&\left(\begin{array}{c}n'\\\vdots\\n'\\k'\\r'\end{array}\right)=(\mathcal{G}_2(N))^T\cdot \left(\begin{array}{c}n\\\vdots\\n\\k\\r\end{array}\right)=\left(\begin{array}{c}n+k+Nr\\\vdots\\n+k+Nr\\-k-2Nr\\r\end{array}\right)&&\Longrightarrow&&\begin{array}{l}k'\,^2-4Nn'r'=k^2-4Nnr \\ k+k'=-2Nr=0\text{ mod 2N}\end{array}\nonumber\\[10pt]
&\left(\begin{array}{c}n'\\\vdots\\n'\\k'\\r'\end{array}\right)=(\mathcal{G}'_2(N))^T\cdot \left(\begin{array}{c}n\\\vdots\\n\\k\\r\end{array}\right)=\left(\begin{array}{c}n\\\vdots\\n\\-2Nn-k\\Nn+k+r\end{array}\right)&&\Longrightarrow&&\begin{array}{l}k'\,^2-4Nn'r'=k^2-4Nnr \\ k+k'=-2Nn=0\text{ mod 2N}\end{array}\nonumber
\end{align}}
\subsubsection{Relation Among Fourier Coefficients of Shift Terms}
In Section~\ref{Sect:ReviewDihedral} we have been careful to point out that $G_{(s_1,s_2)}^{(n,N)}(R,S)$ and $H_{(s_1,s_2)}^{(r,N)}(\rho,S)$ are quasi-Jacobi forms. In particular, the expansions (\ref{ExpansionG}) and (\ref{ExpansionH}) contain the Eisenstein series $E_2$, which under modular transformations produces a non-trivial shift-term (see appendix~\ref{App:ModularStuff}) for details. However, $E_2$ can be promoted to the non-holomorphic modular form $\widehat{E}_2$ of weight $2$ defined in (\ref{NonholEisenstein}). Replacing all $E_2$ in (\ref{ExpansionG}) we can define the following objects
\begin{align}
\widetilde{G}_{(s_1,s_2)}^{(n,N)}(R,S)=\sum_{\ell=0}^{N-1+s_1+s_2}\left(\frac{\pi}{3\,\text{Im} R}\right)^{\ell}\,\widetilde{G}_{(s_1,s_2;\ell)}^{(n,N)}(R,S)\,,
\end{align}
The $\widetilde{G}_{(s_1,s_2;\ell)}^{(n,N)}(R,S)$ are quasi-Jacobi forms of index $Nn$ and weight $s_1+s_2-(2+\ell)$, with $\widetilde{G}_{(s_1,s_2;\ell=0)}^{(n,N)}(R,S)=G_{(s_1,s_2)}^{(n,N)}(R,S)$. These can be expanded in the same fashion as (\ref{DefinitionG})
\begin{align}
\widetilde{G}_{(s_1,s_2;\ell)}^{(n,N)}(R,S)&=\sum_{r=0}^\infty\sum_{k\in \mathbb{Z}}f^{(s_1,s_2;\ell)}_{\underbrace{\text{\scriptsize $n,\ldots,n$}}_{N\text{ times}},k,r}\, Q_S^kQ_R^r\,, &&\forall n\in\mathbb{N}\,,
\end{align}
for $f^{(s_1,s_2;\ell)}_{n,\ldots,n,k,r}\in\mathbb{Z}$. Following the same logic as in the previous Subsection, one would expect 
\begin{align}
\widetilde{G}_{(s_1,0;\ell)}^{(n,N)}(R,S)=\sum_{r=0}^\infty\sum_{k\in\mathbb{Z}}f^{(s_1,0;\ell)}_{\underbrace{\text{\scriptsize $n,\ldots,n$}}_{N\text{ times}},k,r}\,Q_R^r\,Q_S^k=\sum_{r=0}^\infty\sum_{k\in\mathbb{Z}}f^{(s_1,0;\ell)}_{\underbrace{\text{\scriptsize $1,\ldots,1$}}_{N\text{ times}},k,nr}\,Q_R^r\,Q_S^k\,.
\end{align}
which implies
\begin{align}
&(r')^\ell\,f^{(s_1,0;\ell)}_{\underbrace{\text{\scriptsize $n,\ldots,n$}}_{N\text{ times}},k,r}=r^\ell\,f^{(s_1,0;\ell)}_{\underbrace{\text{\scriptsize $n',\ldots,n'$}}_{N\text{ times}},k',r'}\,,\label{RelationsInitialFourier}
\end{align} 
for generic $\ell\in[0,N-1+s_1]$ and
\begin{align}
\left\{\begin{array}{llc}k^2-4Nnr=k'\,^2-4Nn'r'\text{  and  } k=\pm k'\text{ mod }2N & \text{if} & N\notin \mathbb{N}_{\text{prime}}\,, \\ k^2-4Nnr=k'\,^2-4Nn'r' & \text{if} & N\in \mathbb{N}_{\text{prime}}\,.\end{array}\right.
\end{align}
We have systematically verified these conditions up to $N=6$ and $(n,r)\leq (20,4)$. Notice, that it does not matter whether the Fourier coefficients with $\ell>0$ are defined using the modular completion of $G_{(s_1,0;\ell)}^{(n,N)}(R,S)$ (as we did here) or $H_{(s_1,s_2)}^{(r,N)}(\rho,S)$, since in the NS-limit, we have the exchange symmetry $\rho\longleftrightarrow R$. For generic values of $\epsilon_2$ (and thus $s_2$) this symmetry does not exist and thus also (\ref{RelationsGenFourier}) does not hold.
\section{Constraints on $W(R,S,\epsilon_{1,2})$ and Reconstruction}\label{Sect:Reconstruction}
In the second part of this paper, we use the symmetry group $\mathbb{G}(N)$ to argue that there are strong constraints on the function $W(R,S,\epsilon_{1,2})$ appearing in (\ref{ReductionEqW}) such that its explicit form (\ref{FormExplicitW}) seems fixed (at least to leading orders in an expansion of $\epsilon_{1,2}$). To this end, we shall assume that (\ref{ReductionEqW}) holds for $N=2$ for an unspecified function $\unk(R,S,\epsilon_{1,2})$, \emph{i.e.}
\begin{align}
\sum_{s_1,s_2=0}^\infty \epsilon_1^{s_1-1}\epsilon_2^{s_2-1}\,G_{(s_1,s_2)}^{(1,2)}(R,S)=&\,\unk(R,S,\epsilon_{1,2})\sum_{s_1,s_2=0}^\infty \epsilon_1^{s_1-1}\epsilon_2^{s_2-1}\left[G_{(s_1,s_2)}^{(1,1)}(R,S)+K_{(s_1,s_2)}(R,S)\right]\,.\label{ReductionEqTest}
\end{align}
By imposing various constraints on the right hand side of (\ref{ReductionEqTest}), we shall show to order 7 in $\epsilon_{1,2}$ that in fact $\unk(R,S,\epsilon_{1,2})= W(R,S,\epsilon_{1,2})$.
We begin by parametrising a series expansion of $\unk(R,S,\epsilon_{1,2})$ in powers of $\epsilon_{1,2}$
\begin{align}
\unk(R,S,\epsilon_{1,2})=\sum_{s_1,s_2=0}^\infty \epsilon_1^{s_1}\epsilon_2^{s_2} \,\unc_{(s_1,s_2)}(R,S)\,.\label{SeriesUnc}
\end{align}
Notice, since both $G_{(s_1,s_2)}^{(1,2)}(R,S)$ as well as $G_{(s_1,s_2)}^{(1,1)}(R,S)$ in (\ref{ReductionEqTest}) only have a pole of the form $(\epsilon_1\epsilon_2)^{-1}$, the function $\mathfrak{w}(R,S,\epsilon_{1,2})$ needs to be regular for $\epsilon_{1,2}\longrightarrow 0$. Furthermore, to fit the modular properties of $G_{(s_1,s_2)}^{(1,2)}(R,S)$ and $G_{(s_1,s_2)}^{(1,1)}(R,S)$, the coefficients (\ref{SeriesUnc}) are quasi-Jacobi forms of weight $s_1+s_2$ and index $1$ with
\begin{align}
\unc_{(s_1,s_2)}(R,S)&=0\,,&&\forall s_1+s_2\in\mathbb{N}_{\text{odd}}\,,\nonumber\\
\unc_{(s_1,s_2)}(R,S)&=\unc_{(s_2,s_1)}(R,S)\,,&&\forall s_1\,s_2\in\mathbb{N}\,.
\end{align}
Thus similar to eq.~(\ref{ExpansionG}), we can write 
\begin{align}
\unc_{(s_1,s_2)}(R,S)=\sum_{u=0}^{1}w_u^{(s_1,s_2)}(R)\,(\phi_{0,1}(R,S))^{1-u}\,(\phi_{-2,1}(R,S))^{u}\,,\label{ExpansionwL}
\end{align}
where $w_u^{(s_1,s_2)}(R)$ is a quasi-modular form of weight $s_1+s_2+2u$ that can be written as a finite polynomial in the Eisenstein series $\{E_2(R),E_4(R),E_6(R)\}$. Due to its modular properties, the latter can be parametrised by \emph{finitely} many coefficients $c_{u,a}^{(s_1,s_2)}\in\mathbb{Q}$, where the index $a$ runs over a finite set of values. For example, to leading order we can write 
\begin{align}
&w_0^{(0,0)}=c_{0,1}^{(0,0)}\,,&&w_1^{(0,0)}=c_{1,1}^{(0,0)}\,E_2(R)\,,\nonumber\\
&w_0^{(2,0)}=w_0^{(0,2)}=c_{0,1}^{(2,0)} E_2(R)\,,&&w_1^{(2,0)}=w_1^{(0,2)}=c_{1,1}^{(2,0)}\,E_4(R)+c_{1,2}^{(2,0)}\,E^2_2(R)\,,\nonumber\\
&\text{etc.}
\end{align}
which we aim to fix by imposing additional constraints related to the symmetries of the right hand side of (\ref{ReductionEqTest}). Indeed, performing a Fourier expansion of the right hand side of (\ref{ReductionEqTest})  ($\forall n\in\mathbb{N}$)
\begin{align}
\unk(R,S,\epsilon_{1,2})\sum_{s_1,s_2=0}^\infty &\epsilon_1^{s_1-1}\epsilon_2^{s_2-1}\left[G_{(s_1,s_2)}^{(1,1)}(R,S)+K_{(s_1,s_2)}(R,S)\right]\nonumber\\
&=\sum_{s_1,s_2=0}^\infty\sum_{r=0}^\infty\sum_{k\in \mathbb{Z}}\epsilon_1^{s_1-1}\epsilon_2^{s_2-1}h^{(s_1,s_2)}_{k,r}\, Q_S^kQ_R^r\,,
\end{align}
the coefficients $h^{(s_1,s_2)}_{k,r}$ are functions of $c_{u,a}^{(s_1,s_2)}$ and need to satisfy the same symmetry properties just as the $f^{(s_1,s_2)}_{1,1,k,r}$. These conditions in turn can be used to constrain (and fix) the $c_{u,a}^{(s_1,s_2)}$. Concretely, we impose the following conditions:
\begin{itemize}
\item \emph{consistent dihedral action}\\
We can consider the action of various symmetry transformations on the indices $(1,1,k,r)$ (which are the labels of the Fourier coefficients appearing in $G_{(s_1,s_2)}^{(1,2)}(R,S)$ on the left hand side of (\ref{ReductionEqTest})), \emph{e.g.}\footnote{There are other actions of similar type, leading to further relations of the type (\ref{ImposeDihedral}). However, since our studies are only concerned with determining $\unc_{(s_1,s_2)}(R,S)$ for low values of $s_{1,2}$, these relations are sufficient.}
\begin{align}
\mathcal{G}_2(2)^T\cdot(1,1,k,r)^T&=(1+k+2r,1+k+2r, - k-4r,  r)^T\,,\nonumber\\
(\mathcal{G}'_2(2)\cdot \mathcal{B}\cdot\mathcal{G}_2(2))^T\cdot(1,1,k,r)^T&=(9+3k+2r,9+3k+2r,-12-5k-4r,  2+k+r)^T\,,\nonumber\\
(\mathcal{G}_2(2)\cdot \mathcal{B}\cdot\mathcal{G}_2(2))^T\cdot(1,1,k,r)^T&=(1+2k+8r,1+2k+8r,-k-8r, r)^T\,.
\end{align}
Here $\mathcal{B}=\text{diag}(1,1,-1,1)\in \mathbb{Z}_2$ acts by changing the sign of $k$ (which is a symmetry of $G_{(s_1,s_2)}^{(1,2)}(R,S)$, since the latter has even weight under modular transformations). Since 
\begin{align}
&f^{(s_1,s_2)}_{i_1,i_2,k,r}=0&&\text{for} &&\left\{\begin{array}{l}i_1<0\text{ or} \\ i_2<0\,,\end{array}\right.\,,
\end{align}
we can impose respectively
\begin{align}
&h_{k,r}^{(s_1,s_2)}=0\,,&&\forall s_{1,2}\geq 0 &&\text{and for} &&\left\{\begin{array}{l}1+k+2r<0\,, \text{ or}\\9+3k+2r<0\,, \text{ or}\\ 1+2k+8r<0\,.\end{array}\right.\label{ImposeDihedral}
\end{align}
These equations provide implicit (linear) relations for the coefficients $c_{u,a}^{(s_1,s_2)}$ for generic values of $s_{1,2}$.
\item \emph{Supersymmetry Enhancement}\\
As was pointed out in \cite{Haghighat:2013gba,Haghighat:2013tka} (see also \cite{Bastian:2017jje}), for $S=\tfrac{1}{2}(\epsilon_1-\epsilon_2)$, there is a supersymmetry enhancement in the M5-brane setup described in Section~\ref{Sect:ParaEnhancement}, leading to the fact that the (reduced) free energy becomes trivial, \emph{i.e.} 
\begin{align}
\lim_{S\to \frac{1}{2}(\epsilon_1-\epsilon_2)}F_{N,1}(\widehat{a}_1,\ldots,\widehat{a}_N,S,R;\epsilon_{1,2})=1\,,
\end{align}
which in particular also means 
\begin{align}
\lim_{S\to\frac{1}{2}(\epsilon_1-\epsilon_2)}\sum_{s_1,s_2=0}^\infty\sum_{r=0}^\infty\sum_{k\in \mathbb{Z}}\epsilon_1^{s_1-1}\epsilon_2^{s_2-1}\,h^{(s_1,s_2)}_{k,r}\, Q_S^kQ_R^r=1\,,
\end{align}
thus implicitly imposing further constraints on the coefficients $c_{u,a}^{(s_1,s_2)}$. For the case $\epsilon_1=\epsilon_2$, we can in particular deduce
\begin{align}
\sum_{s_1+s_2=\ell}w_{0}^{(s_1,s_2)}=\left\{\begin{array}{lcl}1 & \text{if} & \ell=0\,, \\ 0 & \text{if} & \ell\geq 1\,.\end{array}\right.
\end{align}
\end{itemize}
Using these relations, we have checked up to order $s_1+s_2=7$ that all $w_u^{(s_1,s_2)}$ are fixed uniquely and agree with the expansion (\ref{FormExplicitW}). For the first few, we find explicitly (for simplicity we do not explicitly write the arguments of all functions)
{\allowdisplaybreaks
\begin{align}
\unc_{(0,0)}&=\frac{1}{24}\left[\phi_{0,1}+2\,E_2\,\phi_{-2,1}\right]\,,\nonumber\\
\unc_{(1,1)}&=0\,,\nonumber\\
\unc_{(2,0)}&=\unc_{(0,2)}=\frac{1}{576}\,\left(E_4-E_2^2\right)\, \phi_{-2,1} \,,\nonumber\\
\unc_{(3,1)}&=\unc_{(1,3)}=0\,,\nonumber\\
\unc_{(2,2)}&=\frac{3}{165888}\left[\left(E_2^2-E_4\right)\,\phi_{0,1}+2  \left(5 E_2^3+3 E_4
   E_2-8 E_6\right)\,\phi_{-2,1}\right]\,,\nonumber\\
\unc_{(4,0)}&=\unc_{(0,4)}=\frac{2}{552960}\left[5
   \left(E_4-E_2^2\right)\, \phi_{0,1}+ \left(5 E_2^3+3 E_4 E_2-8 E_6\right)\,\phi_{-2,1}\right]\,.
\end{align}}
Therefore, at least to order $\mathcal{O}(\epsilon_{1,2}^7)$, the function $W(R,S,\epsilon_{1,2})$ that governs the counting of BPS states at the intersection of a single M5 brane with single M2-branes is completely fixed by symmetries. 
\section{Conclusions}\label{Sect:Conclusions}
In this paper we have revisited the symmetries of a particular subsector of the BPS counting function of a system of parallel M5-branes with M2-branes stretched between them that engineers a little string theory with $\mathcal{N}=2$ supersymmetry. While this sector has been studied earlier in the literature, notably in \cite{Ahmed:2017hfr}, here we have taken a slightly different approach, making use of a recent better understanding \cite{Bastian:2018jlf} of symmetries that act directly on the Fourier coefficients of the free energy. The latter can be understood as a consequence of a web of dual low energy theories engineered from the same M5-M2-brane setup. 

From this perspective, we have argued that the reduced free energy $\mathcal{F}^{N}(R,S,\rho)$ (defined in (\ref{DefFreeEnergyReduced})), transforms covariantly under the paramodular group $\Sigma_N$, which acts in a canonical manner on the parameters $(R,S,\rho)$, when arranged in the period matrix $\Omega_N$ (see (\ref{DefPeriodMatrix})). By studying explicitly some of the expansion coefficients of $\mathcal{F}^{N}$, we have furthermore provided non-trivial evidence that in the NS-limit $\Sigma_N$ is in fact enhanced to $\Sigma_N^*\subset Sp(4,\mathbb{R})$ through a single generator, who acts as the exchange $R\longleftrightarrow \rho$. The latter goes beyond the T-duality of the little string theories engineered by the M5-brane setup, since it acts directly on the coefficients of the reduced free energy in the form of (\ref{RelationFourierCoefficientsFreeEnergy}). T-duality would, in our notation, act by replacing $\mathcal{Z}_{N,1}$ with $\mathcal{Z}_{1,N}$. Physically, the explanation for this enhanced symmetry is through an enhancement of the string U-duality group in the NS-limit that was (in a somewhat different context) previously observed in \cite{Hohenegger:2015btj}. The enhancement of the symmetry to $\Sigma_N^*$ is also in agreement with the observation of \cite{Ahmed:2017hfr} that the BPS states contributing to $\mathcal{F}^{N}(R,S,\rho)$, form a symmetric orbifold CFT.

In the second part of the paper we have analysed constraints of various different symmetries on the modular object $W(R,S,\epsilon_{1,2})$, which is crucial for understanding the intersection of a single M5-brane with a single M2-brane on either side (see \figref{fig:PartM5braneConfig} for a graphical representation). Indeed, $W(R,S,\epsilon_{1,2})$ governs locally the counting of BPS states arising in this particular intersection of M-branes. Exploiting various symmetries (and normalisation) of the free energy, we have shown up to order 7 in an expansion of $\epsilon_{1,2}$ that $W(R,S,\epsilon_{1,2})$ is in fact uniquely fixed. Going to higher order is only limited through our computational abilities and one might in fact expect that $W(R,S,\epsilon_{1,2})$ can be entirely fixed in this fashion. 

The work presented in this paper is only the first part in a larger analysis of the symmetries of the free energy: here we have been concerned with the reduced free energy, which captures only a subsector of all the BPS states of the M5-brane system, albeit a sector that has a very rich structure. Physically, this subsector can be characterised to have no weight with respect to all but the imaginary root of the $\widehat{a}_{N-1}$ gauge algebra. In the companion paper \cite{Companion} we shall extend the discussion to the remaining sectors, which probe more deeply the gauge structure. While it is more difficult to make concise statements about the overarching group that contains the modular transformations, the group $\mathbb{G}(N)$ as well as the generators of $\widehat{a}_{N-1}$, we shall exhibit a class of functions that are a certain generalisation of quasi-Jacobi forms and which are relevant in the expansion of the full free energy $F_{N,1}(\widehat{a}_{1,\ldots,N},S,R;\epsilon_{1,2})$. We shall also exhibit the interplay between the reduced free energy and the remaining sectors with regards to the holomorphicity of the free energy.

\section*{Acknowledgements}
We would like to thank Amer Iqbal and Soo-Jong Rey for many discussions and collaborations on related topics. We are in particular thankful to Amer Iqbal, for reading the manuscript prior to publication. Furthermore, we are grateful to Pietro Longhi for useful exchanges and discussions. SH would like to thank the organisers of the GGI workshop 'String Theory from a Worldsheet Perspective' (Galileo Galilei Institute, May 2019) as well as the CERN TH institute 'Topological String Theory and Related Topics' (CERN, June 2019) for creating a stimulating atmosphere, while part of this work was done.

\appendix
\section{Modular Forms}\label{App:ModularStuff}
In this appendix we collect several notions of modular objects that are used throughout the main body of this paper.
\subsection{Jacobi Forms}
A \emph{weak} (or \emph{meromorphic}) \emph{Jacobi form} of index $m\in\mathbb{Z}$ and weight $w\in\mathbb{Z}$ for a finite index subgroup $\Gamma\subset SL(2,\mathbb{Z})$ is a holomorphic function
\begin{align}
\phi:\,\,\mathbb{H}\times\mathbb{C}&\longrightarrow\mathbb{C}\,\nonumber\\
(\rho,z)&\longmapsto \phi(\rho;z)\,,
\end{align}
with $\mathbb{H}$ the upper half-plane, which satisfies
\begin{align}
\phi\left(\frac{a\rho+b}{c\rho+d};\frac{z}{c\rho+d}\right)&=(c\rho+d)^w\,e^{\frac{2\pi i m c z^2}{c\tau+d}}\,\phi(\tau;z)\,,&&\forall\,\left(\begin{array}{cc}a & b \\ c & d\end{array}\right)\in\Gamma\,,\nonumber\\
\phi(\rho;z+\ell_1 \rho+\ell_2)&=e^{-2\pi i m(\ell_1^2\rho+2\ell_1 z)}\,\phi(\rho;z)\,,&&\forall\,\ell_{1,2}\in\mathbb{N}\,,\label{JacobiFormGen}
\end{align}
and which has a Fourier expansion of the form
\begin{align}
\phi(z,\rho)=\sum_{n= 0}^\infty\sum_{\ell\in\mathbb{Z}}c(n,\ell)\,Q_\rho^n\,e^{2\pi i z \ell}\,.
\end{align}
The coefficients satisfy $c(n,\ell)=(-1)^w c(n,-\ell)$. Furthermore, two coefficients $c(n,k)$ and $c(n',k')$ are identical if \cite{EichlerZagier}
\begin{align}
&k^2-4mn=(k')^2-4mn'&&\text{and} &&k=k'\,(\text{mod}\,2m)\,.\label{CoefficientsRelation}
\end{align}
Notice, if $w$ is even and $m=1$ or $m\in\mathbb{N}_{\text{prime}}$, the second condition is redundant. Two examples of Jacobi forms that we shall encounter frequently throughout this work are
\begin{align}
&\phi_{0,1}(\rho,z)=8\sum_{a=2}^4\frac{\theta_a^2(z;\rho)}{\theta_a^2(0,\rho)}\,,&&\text{and}&&\phi_{-2,1}(\rho,z)=\frac{\theta_1^2(z;\rho)}{\eta^6(\rho)}\,,\label{DefPhiFuncts}
\end{align}
where $\theta_{a=1,2,3,4}(z;\rho)$ are the Jacobi theta functions and $\eta(\rho)$ is the Dedekind eta function. Both $\phi_{0,1}$ and $\phi_{-2,1}$ have index one and weight $0$ and $-2$ respectively.\footnote{We remark that for convenience in certain expressions, the numerical pre-factors in (\ref{DefPhiFuncts}) are slightly different than in the literature.}

Notice, the definition of Jacobi forms that we have given above also extends to certain subgroups of $SL(2,\mathbb{Z})$. A congruence subgroup that we shall encounter in this work is $\Gamma_0(\subg(n))$, where $n\in\mathbb{N}$ with the the following prime factor decomposition
\begin{align}
&n=\prod_{i=1}^\ell p_i^{s_i}\,, &&\text{with} &&\ell\in\mathbb{N} &&\text{and} &&\begin{array}{l}s_{1,\ldots,\ell}\in\mathbb{N} \\ p_i\in\mathbb{N}_{\text{prime}} \end{array}\,,\label{PrimeDecomposition}
\end{align}
and the function $\subg$ is defined as.
\begin{align}
\subg(n)=\prod_{i=1}^\ell p_i\,.\label{PrimeFunct}
\end{align}

Let $J_{w,m}(\Gamma)$ be the space of Jacobi forms of index $m$ and weight $w$. We can then define the \emph{$k$th Hecke operator} (for $k\in\mathbb{N}$) in the following fashion 
\begin{align}
\mathcal{H}_k:\,\,J_{w,m}(\Gamma)&\longrightarrow J_{w,km}(\Gamma)\nonumber\\
f_{w,m}(\rho,z)&\longmapsto \mathcal{H}_k(f_{w,m}(\rho,z)) =k^{w-1}\sum_{{d|k}\atop{b\text{ mod }d}}d^{-w}\,f_{w,m}\left(\frac{k\tau+bd}{d^2},\frac{kz}{d}\right)\,.\label{DefHeckeGeneric}
\end{align}
This definition can also be extended to a multi-variable Jacobi form $f_{w,\vec{r}}(\tau,z)$ of weight $w$ and index vector $\vec{r}$, which is mapped to  
\begin{align}
\mathcal{H}_k(f_{w,\vec{r}}(\tau,z)) =k^{w-1}\sum_{{d|k}\atop{b\text{ mod }d}}d^{-w}\,f_{w,\vec{r}}\left(\frac{k\tau+bd}{d^2},\frac{k\vec{z}}{d}\right)\,,
\end{align}
where the right hand side is a multi-variable Jacobi form of weight $w$ and index vector $k\vec{r}$.

\subsection{Quasi-Jacobi Forms and Eisenstein Series}
Throughout most of this paper, we are dealing with quasi-Jacobi forms \cite{Libgober} rather than actual Jacobi forms. We shall discuss the former in more detail in the companion paper \cite{Companion}, here we content ourself by characterising the objects relevant for this work in the following manner
\begin{align}
\psi(\rho;z)=\sum_{i=0}^Kg^u(\rho)\,\phi_{-2,1}^{K-u}(\rho,z)\,\phi_{0,1}^u(\rho,z)\,,
\end{align}
where $g^u(\rho)$ is a quasi-modular form\footnote{We assume that $g^u(\rho)=0$ when $2(K-u)+w<0$.} \cite{QuasiModular2,QuasiModular} of weight $2(K-u)+w$ (such that $\psi$ has weight $w$ and index $K$), which can be written as a polynomial in the Eisenstein series $E_{2k}(\rho)$ for $k\geq 1$. The latter are defined as follows
\begin{align}
&E_{2k}(\rho)=1-\frac{4k}{B_{2k}}\sum_{n=1}^\infty \sigma_{2k-1}(n)\,Q_\rho^n\,,&&\forall\,k\in\mathbb{N}\,,
\end{align}
where $B_{2k}$ are the Bernoulli numbers. The Eisenstein series $E_{2k}$ for $k>1$ can be rewritten as a polynomial in $E_4$ and $E_6$ alone. Furthermore, $E_2$ is not a modular form, but transforms with a shift-term
\begin{align}
&E_2\left(\frac{a\rho+b}{c\rho+d}\right)=(c\rho+d)^2\, E_2(\rho)-\frac{6i}{\pi}\,\frac{c}{c\rho+d}\,,&&\forall\,\left(\begin{array}{cc}a & b \\ c & d\end{array}\right)\in SL(2,\mathbb{Z})\,. \label{ShiftTerm}
\end{align} 
However, we can define the following non-holomorphic series
\begin{align}
\widehat{E}_2(\rho,\bar{\rho})=E_2(\rho)-\frac{6i}{\pi(\rho-\bar{\rho})}\,,\label{NonholEisenstein}
\end{align}
which transforms with weight $2$ under modular transformations, at the expense of being no longer holomorphic.
\section{Paramodular Group}\label{App:Paramod}
For $N\in \mathbb{N}$ the \emph{paramodular group} of level $N$ \cite{Ulrich} is defined in \cite{Dern} as
\begin{align}
&\Sigma_N=\left\{M\in Sp(4,\mathbb{Q})|D_N^{-1}\cdot M\cdot D_N\in\mathbb{Z}^{4\times 4}\right\}\,,&&\text{with}&&D_N=\left(\begin{array}{cc}1\!\!1_{2\times 2} & 0 \\ 0 & P_N\end{array}\right)\,.\label{DefParaModN}
\end{align}
Here and in the following we use the matrices
\begin{align}
&P_N=\left(\begin{array}{cc}1 & 0 \\ 0 & N\end{array}\right)\,,&&U_N=\frac{1}{\sqrt{N}}\left(\begin{array}{cc}0 & N \\ 1 & 0 \end{array}\right)\,,&&\forall N\in \mathbb{N}\,.\label{Defs22Matrices}
\end{align}
The group $\Sigma_N$ can also be written in the form \cite{HeimKrieg,RobertsSchmidt} 
\begin{align}
\Sigma_N=\left\{\left(\begin{array}{cccc}* & *N & * & * \\ * & * & * & */N \\ * & * N & * & * \\ *N & *N & *N & *\end{array}\right)\in Sp(4,\mathbb{Q})\bigg|*\in\mathbb{Z}\right\}\,.\label{AltPresSigmaN}
\end{align}
We can define a natural group action of $\Sigma_N$ on the period matrix $\Omega_N$ defined in (\ref{DefPeriodMatrix})
\begin{align}
M=\left(\begin{array}{cc}A & B \\ C & D\end{array}\right)\in \Sigma_N:&&\Omega_N\longmapsto \Omega'_N=(A\cdot\Omega_N+B)\cdot (C\cdot\Omega_N+D)^{-1}\,.\label{ParaGroupAction}
\end{align}
Furthermore, the generators of $\Sigma_N$ can be defined as \cite{Kappler,Dern} 
\begin{align}
&J_N=\left(\begin{array}{cc}0 & -P_N^{-1} \\ P_N & 0\end{array}\right)\,,&&S_i=\left(\begin{array}{cc}1\!\!1_{2\times 2} & s_i \\ 0 & 1\!\!1_{2\times 2}\end{array}\right)\,,&&\forall i=1,2,3\,,\label{DefGeneratorsSN}
\end{align}
where
\begin{align}
&s_1=\left(\begin{array}{cc}1 & 0 \\ 0 & 0 \end{array}\right)\,,&&s_2=\left(\begin{array}{cc}0 & 1 \\ 1 & 0 \end{array}\right)\,,&&s_3=\left(\begin{array}{cc}0 & 0 \\ 0 & 1/N \end{array}\right)\,.
\end{align}
The action of these generators on $\Omega_N$ according to (\ref{ParaGroupAction}) is given by
\begin{align}
&J_N:&&\Omega_N\longmapsto \Omega'_N=\left(
\begin{array}{cc}
 \frac{\rho }{S^2 N-\rho  R} & \frac{S}{\rho  R-S^2 N} \\
 \frac{S}{\rho  R-S^2 N} & \frac{R}{S^2 N^2-N \rho  R} \\
\end{array}
\right)\,,&&S_1:&&\Omega_N\longmapsto \Omega'_N=\left(
\begin{array}{cc}
 R+1 & S \\
 S & \rho/N \\
\end{array}
\right)\,,\nonumber\\[10pt]
&S_2:&&\Omega_N\longmapsto \Omega'_N=\left(
\begin{array}{cc}
 R & S+1 \\
 S+1 & \frac{\rho }{N} \\
\end{array}
\right)\,,&&S_3:&&\Omega_N\longmapsto \Omega'_N=\left(
\begin{array}{cc}
 R & S \\
 S & \frac{\rho }{N}+\frac{1}{N} \\
\end{array}
\right)\,,\label{ActionParaPeriod}
\end{align}
The generator $J_N$ (and its action on $\Omega_N$) can be better understood by writing $J_N=S_R\cdot S_\rho$, where we defined
\begin{align}
&S_R=\left(
\begin{array}{cccc}
 0 & 0 & -1 & 0 \\
 0 & 1 & 0 & 0 \\
 1 & 0 & 0 & 0 \\
 0 & 0 & 0 & 1 \\
\end{array}
\right): &&\Omega_N\longmapsto \Omega'_N=\left(
\begin{array}{cc}
 -\frac{1}{R} & \frac{S}{R} \\
 \frac{S}{R} & \frac{\rho }{N}-\frac{S^2}{R} \\
\end{array}
\right)\,,\nonumber\\[10pt]
&S_\rho=\left(
\begin{array}{cccc}
 1 & 0 & 0 & 0 \\
 0 & 0 & 0 & -1/N \\
 0 & 0 & 1 & 0 \\
 0 & N & 0 & 0 \\
\end{array}
\right): &&\Omega_N\longmapsto \Omega'_N=\left(
\begin{array}{cc}
 R-\frac{S^2 N}{\rho } & \frac{S}{\rho } \\
 \frac{S}{\rho } & -\frac{1}{N \rho } \\
\end{array}
\right)\,,\label{ActSl2s}
\end{align} 
which essentially act as (one of the) generators of $SL(2,\mathbb{Z})_R$ and $SL(2,\mathbb{Z})_\rho$ respectively, according to (\ref{SL2Actions}).

For $N>1$, the paramodular group $\Sigma_N$ has the following non-trivial extension in $Sp(4,\mathbb{R})$ \cite{RobertsSchmidt,Kappler,Dern}
\begin{align}
&\Sigma^*_N=\Sigma_N\cup \Sigma_N V_N\subset Sp(4,\mathbb{R})\,,&&\text{with} &&V_N=\left(\begin{array}{cc} U_N & 0 \\ 0 & U_N^T\end{array}\right)\,,\label{DefExtensionParamodular}
\end{align}
where $U_N$ is defined in (\ref{Defs22Matrices}). The action of $V_N$ on $\Omega_N$ is given by
\begin{align}
&V_N:&&\Omega_N\longmapsto\Omega'_N=\left(
\begin{array}{cc}
 \rho  & S \\
 S & R/N \\
\end{array}
\right)\,,\label{ActionExtensionParamodular}
\end{align}
corresponding to the exchange $R\longleftrightarrow \rho$.
\section{Fourier Expansion of the Reduced Free Energy}\label{App:SeriesExpansionsOrbifoldSector}
In this appendix we provide evidence for an invariance of the reduced free energy (\ref{DefFreeEnergyReduced}) under the symmetry transformation (\ref{ActionExtensionParamodular}) in the NS-limit. To this end, we verify (\ref{RelationFourierCoefficientsFreeEnergy}) in various examples, while also showing that in general
\begin{align}
&f^{(s_1,s_2)}_{n,\ldots,n,k,r}\neq f^{(s_1,s_2)}_{r,\ldots,r,k,n}\,,&&\text{for}&&s_2>0\,.
\end{align}
\subsection{Example $N=2$}
For $N=2$, the lowest (and most accessible) coefficients to test eq.~(\ref{RelationFourierCoefficientsFreeEnergy}) are the pairs $f^{(s_1,s_2)}_{1,1,k,2}$, $f^{(s_1,s_2)}_{2,2,k,1}$ and $f^{(s_1,s_2)}_{1,1,k,3}$, $f^{(s_1,s_2)}_{3,3,k,1}$
{\allowdisplaybreaks
\begin{align}
\sum_{s_{1,2}=0}^\infty \epsilon_1^{s_1-1}\epsilon_2^{s_2-1} f^{(s_1,s_2)}_{2,2,0,1}&=\frac{3 q^5+37 q^4+118 q^3+118 q^2+37 q+3}{-i\pi  (q-1) q^2\, \epsilon_2}\nonumber\\
&\hspace{0.8cm}+\frac{3 \left(q^4+8 q^3+12 q^2+8 q+1\right)}{q^2}+O(\epsilon_2)\,,\nonumber\\*
\sum_{s_{1,2}=0}^\infty \epsilon_1^{s_1-1}\epsilon_2^{s_2-1} f^{(s_1,s_2)}_{1,1,0,2}&=\frac{3 q^5+37 q^4+118 q^3+118 q^2+37 q+3}{-i\pi  (q-1) q^2\, \epsilon_2}+O(\epsilon_2)\,,\nonumber\\[10pt]
\sum_{s_{1,2}=0}^\infty  \epsilon_1^{s_1-1}\epsilon_2^{s_2-1}f^{(s_1,s_2)}_{2,2,\pm1,1}&=\frac{9 q^4+58 q^3+107 q^2+58 q+9}{-i\pi  (1-q) q^{3/2}\,\epsilon_2}-\frac{8 (q+1)^3}{q^{3/2}}+O(\epsilon_2)\,,\nonumber\\*
\sum_{s_{1,2}=0}^\infty \epsilon_1^{s_1-1}\epsilon_2^{s_2-1} f^{(s_1,s_2)}_{1,1,\pm1,2}&=\frac{9 q^4+58 q^3+107 q^2+58 q+9}{-i\pi  (1-q) q^{3/2}\,\epsilon_2}\phantom{+\frac{8 (q+1)^3}{q^{3/2}}}\hspace{0.2cm}+O(\epsilon_2)\,,\nonumber\\[10pt]
\sum_{s_{1,2}=0}^\infty \epsilon_1^{s_1-1}\epsilon_2^{s_2-1} f^{(s_1,s_2)}_{2,2,\pm2,1}&=\frac{2 \left(5 q^3+21 q^2+21 q+5\right)}{-i\pi  (q-1) q\,\epsilon_2}+6 q+\frac{6}{q}+8+O(\epsilon_2)\,,\nonumber\\*
\sum_{s_{1,2}=0}^\infty \epsilon_1^{s_1-1}\epsilon_2^{s_2-1} f^{(s_1,s_2)}_{1,1,\pm2,2}&=\frac{2 \left(5 q^3+21 q^2+21 q+5\right)}{-i\pi  (q-1) q\,\epsilon_2}\phantom{-6 q-\frac{6}{q}-8}\hspace{0.2cm}+O(\epsilon_2)\,,\nonumber\\[10pt]
\sum_{s_{1,2}=0}^\infty \epsilon_1^{s_1-1}\epsilon_2^{s_2-1} f^{(s_1,s_2)}_{2,2,\pm3,1}&=\frac{\left(5 q^2+13 q+5\right)}{-i\pi  (1-q) \sqrt{q}\,\epsilon_2}+O(\epsilon_2)\,,\nonumber\\*
\sum_{s_{1,2}=0}^\infty \epsilon_1^{s_1-1}\epsilon_2^{s_2-1} f^{(s_1,s_2)}_{1,1,\pm3,2}&=\frac{\left(5 q^2+13 q+5\right)}{-i\pi  (1-q) \sqrt{q}\,\epsilon_2}+O(\epsilon_2)\,,\nonumber\\[10pt]
\sum_{s_{1,2}=0}^\infty \epsilon_1^{s_1-1}\epsilon_2^{s_2-1} f^{(s_1,s_2)}_{2,2,\pm4,1}&=\frac{(q+1)}{-i\pi  (q-1)\,\epsilon_2}+O(\epsilon_2)\,,\nonumber\\*
\sum_{s_{1,2}=0}^\infty \epsilon_1^{s_1-1}\epsilon_2^{s_2-1} f^{(s_1,s_2)}_{1,1,\pm4,2}&=\frac{(q+1)}{-i\pi  (q-1)\,\epsilon_2}+O(\epsilon_2)\,,
\end{align}}
and furthermore
\begin{align}
&f^{(s_1,s_2)}_{2,2,\pm\ell,1}=0=f^{(s_1,s_2)}_{1,1,\ell,2}\,,&&\forall |\ell|\geq 5\,.
\end{align}
Similarly, we have
{\allowdisplaybreaks
\begin{align}
\sum_{s_{1,2}=0}^\infty \epsilon_1^{s_1-1}\epsilon_2^{s_2-1} f^{(s_1,s_2)}_{3,3,0,1}&=\frac{4  (q+1) \left(q^6+14 q^5+64 q^4+122 q^3+64 q^2+14 q+1\right)}{-i\pi  (q-1) q^3\,\epsilon_2}\nonumber\\
&\hspace{0.8cm}+\frac{8 (q+1)^4 \left(q^2+8 q+1\right)}{q^3}+\mathcal{O}(\epsilon_2)\nonumber\\
\sum_{s_{1,2}=0}^\infty \epsilon_1^{s_1-1}\epsilon_2^{s_2-1} f^{(s_1,s_2)}_{1,1,0,3}&=\frac{4  (q+1) \left(q^6+14 q^5+64 q^4+122 q^3+64 q^2+14 q+1\right)}{-i\pi  (q-1) q^3\,\epsilon_2}+\mathcal{O}(\epsilon_2)\nonumber\\[10pt]
\sum_{s_{1,2}=0}^\infty \epsilon_1^{s_1-1}\epsilon_2^{s_2-1} f^{(s_1,s_2)}_{3,3,\pm1,1}&=\frac{13 q^6+114 q^5+425 q^4+670 q^3+425 q^2+114 q+13}{-i\pi  (1-q) q^{5/2}\,\epsilon_2}\nonumber\\
&\hspace{0.8cm}-\frac{24 (q+1)^3 \left(q^2+3 q+1\right)}{q^{5/2}}+O(\epsilon_2)\,,\nonumber\\*
\sum_{s_{1,2}=0}^\infty \epsilon_1^{s_1-1}\epsilon_2^{s_2-1} f^{(s_1,s_2)}_{1,1,\pm1,3}&=\frac{13 q^6+114 q^5+425 q^4+670 q^3+425 q^2+114 q+13}{-i\pi  (1-q) q^{5/2}\,\epsilon_2}+O(\epsilon_2)\,,\nonumber\\[10pt]
\sum_{s_{1,2}=0}^\infty \epsilon_1^{s_1-1}\epsilon_2^{s_2-1} f^{(s_1,s_2)}_{3,3,\pm2,1}&=\frac{16 (q+1) \left(q^4+6 q^3+13 q^2+6 q+1\right)}{-i\pi  (q-1) q^2\,\epsilon_2}+\frac{24 (q+1)^4}{q^2}+O(\epsilon_2)\,,\nonumber\\*
\sum_{s_{1,2}=0}^\infty \epsilon_1^{s_1-1}\epsilon_2^{s_2-1} f^{(s_1,s_2)}_{1,1,\pm2,3}&=\frac{16 (q+1) \left(q^4+6 q^3+13 q^2+6 q+1\right)}{-i\pi  (q-1) q^2\,\epsilon_2}\phantom{-\frac{24 (q+1)^4}{q^2}}\hspace{0.2cm}+O(\epsilon_2)\,,\nonumber\\[10pt]
\sum_{s_{1,2}=0}^\infty \epsilon_1^{s_1-1}\epsilon_2^{s_2-1} f^{(s_1,s_2)}_{3,3,\pm3,1}&=\frac{\left(9 q^4+58 q^3+107 q^2+58 q+9\right)}{-i\pi  (1-q) q^{3/2}\,\epsilon_2}-\frac{8 (q+1)^3}{q^{3/2}}+O(\epsilon_2)\,,\nonumber\\*
\sum_{s_{1,2}=0}^\infty \epsilon_1^{s_1-1}\epsilon_2^{s_2-1} f^{(s_1,s_2)}_{1,1,\pm3,3}&=\frac{\left(9 q^4+58 q^3+107 q^2+58 q+9\right)}{-i\pi  (1-q) q^{3/2}\,\epsilon_2}+\phantom{\frac{8 (q+1)^3}{q^{3/2}}}\hspace{0.2cm}+O(\epsilon_2)\,,\nonumber\\[10pt]
\sum_{s_{1,2}=0}^\infty \epsilon_1^{s_1-1}\epsilon_2^{s_2-1} f^{(s_1,s_2)}_{3,3,\pm4,1}&=\frac{2  \left(q^3+7 q^2+7 q+1\right)}{-i \pi  (q-1) q\,\epsilon_2}+O(\epsilon_2)\,,\nonumber\\*
\sum_{s_{1,2}=0}^\infty \epsilon_1^{s_1-1}\epsilon_2^{s_2-1} f^{(s_1,s_2)}_{1,1,\pm4,3}&=\frac{2  \left(q^3+7 q^2+7 q+1\right)}{-i \pi  (q-1) q\,\epsilon_2}+O(\epsilon_2)\,,\nonumber\\[10pt]
\sum_{s_{1,2}=0}^\infty \epsilon_1^{s_1-1}\epsilon_2^{s_2-1} f^{(s_1,s_2)}_{3,3,\pm5,1}&=\frac{\sqrt{q}}{-i\pi  (1-q)\,\epsilon_2}+O(\epsilon_2)\,,\nonumber\\*
\sum_{s_{1,2}=0}^\infty \epsilon_1^{s_1-1}\epsilon_2^{s_2-1} f^{(s_1,s_2)}_{1,1,\pm5,3}&=\frac{\sqrt{q}}{-i\pi  (1-q)\,\epsilon_2}+O(\epsilon_2)\,,
\end{align}}
and furthermore
\begin{align}
&f^{(s_1,s_2)}_{3,3,\pm\ell,1}=0=f^{(s_1,s_2)}_{1,1,\ell,3}\,,&&\forall |\ell|\geq 6\,.
\end{align}
This implies
\begin{align}
&f^{(s_1,0)}_{2,2,k,1}=f^{(s_1,0)}_{1,1,k,2}\,,&&k\in\mathbb{Z}\,,&&s_1\geq 0\,,\nonumber\\
&f^{(s_1,0)}_{3,3,k,1}=f^{(s_1,0)}_{1,1,k,3}\,,&&k\in\mathbb{Z}\,,&&s_1\geq 0\,.
\end{align}
thus confirming (\ref{RelationFourierCoefficientsFreeEnergy}), as well as the existences of coefficients with
\begin{align}
&f^{(s_1,s_2)}_{n,n,k,r}\neq f^{(s_1,0)}_{r,r,k,n}\,,&&\text{for}&&s_1\geq 0\,,&&s_2\geq 1\,.
\end{align}
Notice, that (due to the symmetry properties of $W(R,S,\epsilon_{1,2})$ in (\ref{FormExplicitW}) as well as (\ref{ReductionEqW})), $\sum_{s_{1,2}=0}^\infty \epsilon_1^{s_1-1}\epsilon_2^{s_2-1} f^{(s_1,s_2)}_{1,1,k,n}$ is an odd function in $\epsilon_2$, which explains the absence of the corresponding terms in the above expansions.

Finally, focusing on the case $s_1=0=s_2$, we can supplement further evidence for eq.~(\ref{RelationFourierCoefficientsFreeEnergy})
\begin{align}
\sum_{k\in\mathbb{Z}}&f^{(0,0)}_{2,2,k,3}\,Q_S^k=\sum_{k\in\mathbb{Z}}f^{(0,0)}_{3,3,k,2}\,Q_S^k=2 \left(Q_s^7+\frac{1}{Q_s^7}\right)-208 \left(Q_s^6+\frac{1}{Q_s^6}\right)+3548\left(Q_s^5+\frac{1}{Q_s^5}\right)\nonumber\\
&-25840\left(Q_s^4+\frac{1}{Q_s^4}\right)+105592 \left(Q_s^3+\frac{1}{Q_s^3}\right)-272752 \left(Q_s^2+\frac{1}{Q_s^2}\right)+472810\left(Q_s+\frac{1}{Q_s}\right)-566304\,.
\end{align}
\subsection{Example $N\geq 3$}
For the cases $N\geq 3$, we provide further evidence for eq.~(\ref{RelationFourierCoefficientsFreeEnergy}) for $s_1=0=s_2$.
\begin{align}
\sum_{k\in\mathbb{Z}}&f^{(0,0)}_{2,2,2,k,1}\,Q_S^k=\sum_{k\in\mathbb{Z}}f^{(0,0)}_{1,1,1,k,2}\,Q_S^k=3 \left(Q_s^5+\frac{1}{Q_s^5}\right)-42 \left(Q_s^4+\frac{1}{Q_s^4}\right)+270\left(Q_s^3+\frac{1}{Q_s^3}\right)\nonumber\\
&-948 \left(Q_s^2+\frac{1}{Q_s^2}\right)+1959\left(Q_s+\frac{1}{Q_s}\right)-2484\,,\nonumber\\[10pt]
\sum_{k\in\mathbb{Z}}&f^{(0,0)}_{3,3,3,k,1}\,Q_S^k=\sum_{k\in\mathbb{Z}}f^{(0,0)}_{1,1,1,k,3}\,Q_S^k=-6 \left(Q_s^6+\frac{1}{Q_s^6}\right)+114 \left(Q_s^5+\frac{1}{Q_s^5}\right)-948\left(Q_s^4+\frac{1}{Q_s^4}\right)\nonumber\\
&+4362 \left(Q_s^3+\frac{1}{Q_s^3}\right)-12306\left(Q_s^2+\frac{1}{Q_s^2}\right)+22500 \left(Q_s+\frac{1}{Q_s}\right)-27432
\end{align}
\begin{align}
\sum_{k\in\mathbb{Z}}&f^{(0,0)}_{2,2,2,2,k,1}\,Q_S^k=\sum_{k\in\mathbb{Z}}f^{(0,0)}_{1,1,1,1,k,2}\,Q_S^k=12 \left(Q_s^5+\frac{1}{Q_s^5}\right)-144 \left(Q_s^4+\frac{1}{Q_s^4}\right)+808\left(Q_s^3+\frac{1}{Q_s^3}\right)\nonumber\\
&-2592 \left(Q_s^2+\frac{1}{Q_s^2}\right)+5100\left(Q_s+\frac{1}{Q_s}\right)-6368\,.
\end{align}

\end{document}